\def\year{2022}\relax
\documentclass[letterpaper]{article}
\usepackage{aaai22} 
\usepackage{times} 
\usepackage{helvet} 
\usepackage{courier} 
\usepackage[hyphens]{url} 
\usepackage{graphicx} 
\usepackage{graphicx} 
\usepackage{natbib} 
\usepackage{caption} 
\usepackage{hyperref}
\usepackage{array}  
\usepackage{multirow}
\usepackage{booktabs}
\usepackage{subcaption}
\usepackage{graphicx}
 \usepackage{float}
\usepackage{nameref}
\usepackage{varioref}
\usepackage{hyperref}
\hypersetup{
    colorlinks,
    linkcolor={red!50!black},
    citecolor={blue!50!black},
    urlcolor={blue!80!black}
}
\usepackage{cleveref}

\DeclareCaptionStyle{ruled}%
{labelfont=normalfont,labelsep=colon,strut=off}
\usepackage{xcolor}

\title{Comparing Methods for Creating a National Random Sample of Twitter Users}

\author{Meysam Alizadeh\thanks{\scriptsize{Department of Political Science, University of Zurich. Email: \texttt{\href{mailto:alizadeh@ipz.uzh.ch}{alizadeh@ipz.uzh.ch}.} Google Scholar: \href{https://scholar.google.com/citations?user=FyKst9AAAAAJ&hl=en}{https://scholar.google.com/citations?user=FyKst9AAAAAJ\&hl=en}. Corresponding Author.}}, Darya Zare\thanks{\scriptsize{Department of Computer Engineering, Amirkabir University of Technology, Tehran, Iran.}}, Zeynab Samei\thanks{\scriptsize{Department of Computer Science, Institute For Research In Fundamental Sciences, Tehran, Iran.}}, Mohammadamin Alizadeh \thanks{\scriptsize{Department of Mathematics and Computer Science, Amirkabir University of Technology.}}, Mael Kubli\thanks{\scriptsize{Department of Political Science, University of Zurich.}}, \\ Mohammadhadi Aliahmadi\thanks{\scriptsize{Department of Industrial Engineering, Iran University of Science and Technology, Tehran, Iran.}}, Sarvenaz Ebrahimi\thanks{\scriptsize{Faculty of Entrepreneurship, University of Tehran, Tehran, Iran.}}, Fabrizio Gilardi\thanks{\scriptsize{Department of Political Science, University of Zurich.}} }

\begin{document}

\maketitle
\begin{abstract}
Twitter data has been widely used by researchers across various social and computer science disciplines. A common aim when working with Twitter data is the construction of a random sample of users from a given country. However, while several methods have been proposed in the literature, their comparative performance is mostly unexplored. In this paper, we implement four common methods to collect a random sample of Twitter users in the US: \textit{1\% Stream}, \textit{Bounding Box}, \textit{Location Query}, and \textit{Language Query}. Then, we compare the methods according to their tweet- and user-level metrics as well as their accuracy in estimating US population with and without using inclusion probabilities of various demographics. Our results show that the \textit{1\% Stream} method performs differently than others in tweet- and user-level metrics, and best for the construction of a population representative sample. We discuss the conditions under which the \textit{1\% Stream} method may not be suitable and suggest the \textit{Bounding Box} method as the second-best method to use.
\end{abstract}



\section{Introduction}

Twitter data have been widely used by researchers across various social and computer sciences disciplines \cite{king2017computer}. One of the key challenges in working with Twitter data is to obtain a random sample of users from a country \cite{kim2018evaluating}. The goal is usually to get a platform or population-representative sample of users \cite{wang2019demographic}. The sample is then used for public opinion research, for experimental research, or for training machine learning algorithms. For example, random samples of Twitter users have been used for estimating public opinion \cite{barbera2019leads, alizadeh2019psychology, alizadeh2014distributions}, studying the diffusion of misinformation \cite{shao2018anatomy}, studying conspiracy theories \cite{batzdorfer2022conspiracy}, evaluating the performance of large language models for text annotation tasks \cite{alizadeh2023open}, measuring the influence of information operations \cite{barrie2021kingdom}, and developing supervised models for detecting inauthentic activities \cite{alizadeh2020content}. However, there are at least two significant challenges in obtaining a random sample of users from a country: 1) while several methods have been proposed in the literature, it is not clear which one is the best, and 2) the extent to which these random samples are actually representative of the population is questionable.

There are at least four popular methods to construct a random sample of Twitter users for a specific country. First, at the moment of writing, Twitter provides 1\% of all tweets worldwide in real-time through its free stream API. One can collect this stream for a specific period of time, even with filtering for language or country of interest, obtain a list of users who posted tweets, and then sample from them. Second, it is possible to use Twitter's search-tweet API and query for a specific language, and after ingesting tweets for a period of time, filter for a language of interest, which matches different countries to different degrees. Third, one can directly query for a country of interest using the Search API. Fourth, the Twitter API allows queries based on the bounding box coordinates enclosing a specific country and researchers used it to get a random sample of a country's Twitter users (e.g. \cite{barrie2021kingdom}). 

The extent to which these four methods produce similar results, and which one produces a more representative sample of a population, is mostly unexplored. Although Twitter shut off the free access to the API in February 2023, many researchers have archived a wealth of data and still publishing novel research using Twitter data (e.g. \cite{truong2024account, mosleh_rand_2024}). Moreover, in compliance with the EU's Digital Service Act, Twitter is now accepting researcher API access applications. In this paper, we compare these methods with respect to fourteen metrics at the level of tweets (e.g. distribution of the number of tweets per day), users (e.g. distribution of age and gender), and population. For the population-level metrics, our goal is to investigate which of the four Twitter sampling methods provides the best data for creating a nationally representative sample of users. To this end, we follow the approach proposed in \cite{wang2019demographic} and create representative samples from each of the four Twitter sampling methods and use them separately to estimate a population of US from Twitter data.

In the following text, we review the theoretical background and related work to highlight the need and importance of comparing various existing Twitter sampling methods. Next, we discuss our methodology to collect Twitter users, exclude non-active, non-individual, and bot-like accounts, infer the age, gender, and location, create nationally-representative samples from them, and compare them according to various evaluation metrics, which we devised from carefully reviewing the literature. 

The results show that the \textit{1\%Stream} Twitter sampling method is the one that produces the best population-representative sample and which exhibits different characteristics, compared to the other three sampling methods. The results further underscore the \textit{Bounding Box} sampling method as the best replacement for situations under which the \textit{1\% Stream} method might not be feasible or suitable. Our results illuminate the positive and negative characteristics of each sampling method and help researchers choose the one that best suits their research goals and designs. By identifying the best sampling methods, our results also pave the way for conducting more accurate social listening studies and building more accurate machine learning models.

\section{Related Work}

Generally speaking, there are two types of methodological approaches to collecting social media data, including Twitter: 1) 
\textit{keyword-based} approach, and 2) \textit{sample-based} approach. In the keyword-based approach, researchers create a list of hashtags or keywords and collect all matched tweets over a period of time. Although this approach is popular due to its ease of automation, it suffers from some shortcomings (see \cite{kim2018evaluating} for a discussion). The most crucial drawback of the keyword-based data collection is that most researchers often pick keywords in ad hoc ways that are far from optimal and usually biased (see \cite{king2017computer} and \cite{munger2022political} for potential solutions). Other researchers take the sample-based approach, which is the focus of this study, and try to sample desired tweets or users.

From the tweet sampling perspective, since Twitter has yet to be transparent about how its data sampling is performed, early research on getting data from Twitter data were focused on understanding its underlying mechanisms. Comparing the limited Streaming API with the unlimited but costly Firehose API, \cite{morstatter2013sample} tried to answer whether data obtained through Twitter’s sampled Streaming API is a sufficient representation of activity on Twitter as a whole. They found that for larger numbers of matched tweets, the Streaming API's coverage is reduced, but its ability to estimate the top hashtags is comparable to the Firehose. Interestingly, the results showed that the Streaming API almost returns the complete set of geotagged tweets despite sampling. 

One of the critical questions about the Twitter Streaming API is whether it provides similar results to that of multiple simultaneous API requests from different connections. \cite{joseph2014two} compared samples of tweets collected using the Streaming API that tracked the same set of keywords at the same time. Their results showed that, on average, over 96\% of the tweets in various samples are the same. In practice, this means that an infinite number of Streaming API samples are required to collect most of the tweets containing a particular popular keyword \cite{joseph2014two}. In another work, \cite{tufekci2014big} proposed a framework to address potential biases in tweet collection and how to mitigate them.

More recently, \cite{kim2018evaluating} compared simple daily random sampling with constructed weekly sampling. Their results underscored simple daily random sampling as the efficient way to obtain a representative sample of tweets for a specific period of time. In another important study, \cite{pfeffer2018tampering} showed that the Streaming API can be deliberately manipulated by adversarial actors due to the nature of Twitter's sampling mechanism. Their results further showed that technical artifacts of the Streaming API can skew tweet samples and therefore those samples should not be regarded as random.    

In a distinct investigation, researchers had reverse-engineered the sampling mechanism used by Twitter's Sample API. The sample was based on the timestamp of when tweets arrived at Twitter's servers and any tweet arriving between 657-666 milliseconds was included in the 1\% Sample API \cite{pfeffer2018tampering}.

In \cite{de2010does}, researchers explored the impact of attribute and topology-based sampling strategies on the discovery of information diffusion in Twitter. The study analyzed several widely-adopted sampling methods that selected nodes based on attributes and topology, and developed metrics based on user activity, topology, and temporal characteristics to evaluate the sample's quality. The results showed that incorporating both network topology and user-contextual attributes significantly improved the estimation of information diffusion by 15-20\%.

Another research highlighted the lack of common standards for data collection and sampling in the emerging field of digital media and social interactions. The paper focused on Twitter and compared the networks of communication reconstructed using different sampling strategies. The paper concluded that a more careful account of data quality and bias, and the creation of standards that can facilitate the comparability of findings, would benefit the emerging area of research \cite{gonzalez2014assessing}.

In another work, \cite{wang2015should} compared the representativeness of two Twitter data samples obtained from the Twitter stream API, Spritzer and Gardenhose, with a more complete Twitter dataset. The study found that both sample datasets capture the daily and hourly activity patterns of Twitter users and provide representative samples of the public tweets, but tend to overestimate the proportion of low-frequency users.

%
%

A detailed analysis of the effects of Twitter data sampling on measurement and modeling studies across different timescales and subjects was presented in \cite{wu2020variation}. It validated the accuracy of Twitter rate limit messages in approximating the volume of missing tweets and identified significant temporal and structural variations in the sampling rates across different scales and entities. They also suggested the use of the Bernoulli process with a uniform rate for counting statistics and provided effective methods for estimating ground-truth statistics. 

This paper \cite{hino2019representing} addressed the challenges researchers face in accessing representative and high-quality data from social media platforms like Twitter. The authors proposed a methodology for creating a cost-effective and accessible archive of Twitter data through population sampling, resulting in a highly representative database. The study demonstrated the high degree of representativeness achieved by comparing the sample data with the ground truth of Twitter's full data feed, making it suitable for post-hoc analyses and enabling researchers to refine their keyword searches and collection strategies. Overall, this approach provided an alternative solution for researchers with limited resources to access social media data under resource constraints.

In terms of comparing expert sampling and random sampling,
\cite{zafar2015sampling} explored the advantages and disadvantages of the two methods. The study found that expert sampling offers a number of advantages over random sampling, including more rich information content, trustworthiness, and timely capture of important news and events. However, random sampling preserves the statistical properties of the entire data set and automatically adapts to the growth and changes of the network, while expert sampling does not. The authors suggested that both random and expert sampling techniques would be needed in the future, and called for equal focus on expert sampling of social network data.

From the Twitter user sampling perspective, the most important issue that researchers have tried to tackle is the problem of sampling bias. Indeed, much of the extant literature on sampling users from Twitter is related to pointing out sampling biases in election prediction studies and the necessity to control for it (see \cite{jungherr2012pirate} for a discussion and \cite{gayo2013meta} for an early review). More recently, \cite{yang2022twitter} focused on the issue of inauthentic accounts that could skew the behavior of voters. They proposed a method to identify potential voters on Twitter and compared their behavior with various samples of American Twitter users. The results showed that users sampled from the Streaming API are more active and conservative compared to the potential voters and randomly selected users. They further showed that the users in the Streaming API sample tend to exhibit more inauthentic behaviors, involve in more bot-like activities, and share more links to low-credibility sources.

Although the problem of sampling bias in Twitter has been recognized in many works, none of the papers we discussed above addressed the issue, due to the lack of a valid methodology to do so. In a seminal work, \cite{wang2019demographic} proposed a method by combining demographic inference with post-stratification to make social media samples a more representative of a population. First, they created a multimodal deep neural network classifier for joint identification of age, gender, and non-individual accounts. Second, they proposed a multilevel logistic regression approach to correct for sampling biases. The proposed debiasing approach estimates inclusion probabilities of users from various demographic groups from inferred joint populations and ground-truth population histograms. They further showed that their fully debiased sample outperforms a baseline and marginally debiased samples in the prediction task of estimating European regions' population from Twitter data.

Creating a random sample of Twitter users from a country or geographical region has been cited in some of the research discussed above (e.g. \cite{pfeffer2018tampering, yang2022twitter}). However, no study pointed out the fact that there are several methods for creating such a sample from Twitter users. For example, one can use Streaming API to collect tweets published in a certain country or language and then randomly sample from the list of tweets' authors, or just simply use the Search API and query for country or language or both. We identified four such Twitter sampling methods in the literature (plus a fifth one which is not feasible anymore). The extent to which these random (or near-random) sampling methods produce similar results or are more representative of a population is unexplored. In this paper, we attempt to provide answers to the following research questions:

\begin{itemize}
    \item RQ1: Do different methods for creating a random sample of Twitter users from a country produce similar results in terms of the tweet- and user-level metrics?
    \item RQ2: Are different methods for creating a random sample of Twitter users from a country produce representative sample of the population? If not, which method provides a more representative sample of the population?
\end{itemize}

To answer these research questions, we collect US Twitter data using four widely-used Twitter user sampling methods for one month and compare the results according to multiple tweet- (e.g. distribution of tweets), user- (e.g. distribution of age and gender), and population-level evaluation metrics. For our population-level metrics, following \cite{wang2019demographic}, we use five different prediction errors for estimating the population of US from Twitter data. The goal of the population-level metrics is to explore which sampling and debiasing methods provide the minimum prediction error, and thus, more representative of the population.

\section{Methodology}
 
\subsection{Sampling Methods}
We use four different Twitter sampling methods (Table \ref{tab:methods}) that are widely used in the literature to create a random sample of Twitter users in a country. They include 1) \textit{1\% Stream}, in which we use Twitter Streaming API to get 1\% stream tweets, and filter for languages or country of interest; 2) \textit{Country Query}, in which we query for the country into Twitter Search API and get all tweets; 3) \textit{Language Query}, in which we query for languages that are related to the country of interest and use Twitter Search API same as \textit{Country Query}; and 4) \textit{Bounding Box}, in which we divide a country to multiple small bounding-box coordinates and get all tweets within them by Twitter Search API. There is also a fifth sampling method, in which one could generate random Twitter user IDs, check whether it exists on Twitter, and if they existed, filter to country or language of interest \cite{barbera2019leads}. However, this method is no longer feasible due to technical changes on Twitter \footnote{For a comprehensive explanation and description of our improved method, please refer to Appendix \nameref{sec:appendix_data_collection}.}.

\begin{table*}[t]
\centering 
\footnotesize
\begin{tabular}{m{0.3cm}m{2.6cm}m{2cm}m{11cm}}
\hline \\[-1.8ex] 
No. & Method & Name & Description \\ 
\hline \\[-1.8ex] 
1 & Streaming API & Stream 1\% & Get 1\% stream tweets for a month, filter for potential country field.\\
2 & Country Query & Loc & Query for the country and get all tweets for a month.\\
3 & Language Query  & Lang & Query for relevant languages and the country and get all tweets for a month. \\
4 & Bounding Box & BB & Divide a country to multiple small bounding-box coordinates and get all tweets within them for a month.\\
\hline \\[-1.8ex] 
\end{tabular} 
\caption{\footnotesize Four different methods of getting a random sample of users for a country.}
\label{tab:methods} 
\end{table*} 

\subsection{Data}

We used Twitter's V2 API to collect tweets over a month (i.e. from 2022-09-07 to 2022-10-08). As a first method, we used the sampled Streaming API. This is the simplest method that returns 1 \% of all tweets during the listening period, but it generates the most amount of noise as well. For the second method, we used Twitter Search API and collected tweets posted in the United States by setting the filter argument as \emph{place\_country:US}. For the third method, we used the same endpoint as the second one but filtered tweets based on English language and US country by setting the filter argument as \emph{lang:en} and  \emph{place\_country:US}. As for the fourth method, we used the bounding boxes available in the Twitter Search API, which uses the coordinates of a specific area. The bounding boxes are limited in size (25 miles in height and width) and their form (rectangular). To mitigate the limitations of the bounding boxes, we implemented a grid of small boxes plotted over the US, with points defined by a longitudinal and latitudinal distance of 0.3°. We also included points along the borders of the countries by shifting the points diagonally up and down for one unit. This resulted in a comprehensive approximation of the United States, represented by 9,541 bounding boxes. In summary, Our research scope focuses on Twitter users in the United States. However, we applied certain query parameters in all four methods to narrow down our user selection within this context. It's important to note that we did not use any specific keywords or hashtags that could potentially introduce bias towards any particular topic.


\subsection{User Pre-Processing} \label{sec:prepro}
Following the data collection, the next step involves the pre-processing of accounts. Typically, when we generate a random sample of Twitter users from a specific country, our goal is to obtain a sample consisting of authentic individuals rather than organizations or malicious accounts like bots. This is important as the inclusion of such accounts could introduce bias into our sample  \cite{yang2022twitter}. To achieve this, we selected a random sample of 30K users from each dataset associated with each of the four Twitter sampling methods. This equalizes the dataset sizes and facilitates a more accurate comparison of changes in data volume after applying each pre-processing filter. From the initial sample of 30K randomly selected users, we applied several filters to exclude specific categories, including bots (identified using botometer \cite{yang2022botometer}), verified accounts, protected accounts, low-activity accounts (those with fewer than 100 tweets), recently created accounts (less than 9 months old), and suspended accounts (as detailed in Table \ref{tab:filter methods}). Additionally, we eliminated accounts whose bios contain keywords such as "journalist," "magazine," "member," "organization," "mayor," "actress," etc., as outlined in Table \ref{tab:filter methods}. This step ensures that our sample comprises regular individuals rather than celebrities or organizational accounts. Lastly, we excluded users whose tweet language is not English and those whose tweet coordinates did not correspond to locations in the United States.


\begin{table}[t]
\centering 
\footnotesize
\begin{tabular}{m{0.3cm}l p{5cm}}
\hline \\[-1.8ex] 
No. & Filter & Description \\ 
\hline \\[-1.8ex] 
1 & Verified & Exclude Twitter accounts that have been verified by Twitter.\\
2 & Activity & Exclude users who have posted fewer than 100 tweets during their life time on Twitter.\\
3 & Tenure & Exclude users who have created their accounts within the nine months leading up to the data collection.\\
4 & Biography & Exclude users whose bios contain any of the following terms: journalist, anchor, newspaper, representative, congressman, congresswoman, senator, secretary, mayor, organization, organization, company, institute, charity, magazine, singer, bot, member, advisory, advisor, startup, venture, news, actor, actress, official page.\\
5 & Language & Exclude users from the USA samples whose tweets are not in the English language.\\
6 & Country & Exclude users whose tweets are not geotagged to locations within the United States.\\
7 & Protected & Exclude users whose accounts are set to "protected" status at the time of analysis.\\
8 & Suspended & Exclude users whose accounts are suspended or deleted at the time of analysis.\\
\hline \\[-1.8ex] 
\end{tabular} 
\caption{\footnotesize Pre-Processing Steps for Twitter Users}
\label{tab:filter methods} 
\end{table}

\subsection{Inferring Users' Demographics}\label{sec:debiasing}
We utilized the M3 model \cite{wang2019demographic}, which is a multimodal deep learning model, to predict the gender and age of Twitter users. This model functions across 32 languages and relies solely on users' profile details, including their \textit{screen\_name}, \textit{user\_name}, \textit{bio}, and \textit{profile picture}. The M3 model offers two modes, and we employed the 'full' model, which is more accurate and incorporates the profile image. In addition to gender and age, the model also discerns whether a Twitter account belongs to an organization. Consequently, we excluded organizational accounts from our dataset.
Regarding location, we followed established research practices by utilizing self-reported user location or information from their close connections to estimate the users' locations at the state level \cite{barbera2019leads}.


\subsection{Creating Representative Population Estimates}
In this step, we've inferred all the necessary features, enabling us to create our sample. We've selected all the valid users that have passed through the previous filters, resulting in our final 10K samples for each method. From this point forward, we will use this 10K-sample dataset for our analysis.
Previous research has demonstrated that when demographic information is available, and proper statistical adjustments like re-weighting and post-stratification are applied, non-representative polls can still yield accurate population estimates \cite{wang2015forecasting}. The primary demographic characteristics that survey analysts focus on to address non-representativeness are age, gender, and location \cite{wang2019demographic}. We follow the approach introduced in \cite{wang2019demographic} to learn inclusion probabilities based on users' demographics. This method utilizes multilevel regression techniques to estimate the likelihood of an individual with specific demographics being present on a particular platform. It does this by considering inferred joint population counts and ground-truth population data. To implement this approach, we require gender, age, and geographic location information for each Twitter user, along with ground-truth data about the population. In the case of the United States, we rely on census data as our ground truth.\footnote{\href{https://www2.census.gov/programs-surveys/popest/datasets/2020-2021/state/asrh/}{census.gov}}).





\subsection{Evaluation Metrics}\label{sec:eval}

We compare the results of each method to create a nationally-random sample of Twitter users according to three categories of metrics including 1) tweet-level (Table \ref{tab:evaluation}), 2) account-level (Table \ref{tab:evaluation}), and 3) population-level metrics (Table \ref{tab:evaluation population}). Tweet-level metrics include total number of tweets generated by each sampling method, average number of tweets collected from each account, and share of English tweets. User-level metrics encompass various aspects, including: Total number of unique users, Distribution of tweet counts for accounts, considering the correlation between tweet count and account age, Distribution of the average tweet count, calculated by dividing the tweet count by the account's age in days, Distribution of the number of likes received by the last tweet of each user, Distribution of account creation dates, Distribution of the number of followers and friends by each user, and
Distribution of age and gender among users.

\begin{table*}[t]
\centering 
\footnotesize
\begin{tabular}{m{1.8cm}m{5.5cm}m{7.5cm}}
\hline \\[-1.8ex]
Category & Criteria & Description \\
\hline \\[-1.8ex]
 & Number of tweets &  Total number of collected tweets.\\
Tweet-Level & Average tweet per account & Average number of tweets per account.\\
 & Relevant language & Share of tweets in country-specific languages.\\
\hline \\[-1.8ex] 
& Number of accounts & Number of unique accounts.\\
& Distribution of tweet count& Distribution of total number of tweets for each account.\\
& Distribution of Average tweet count& The Distribution of tweets per day is calculated as the total tweet count for each account, divided by the age of the account.\\

Account-Level & Distribution of likes &  For each account, distributions of the number of last tweets' likes.\\
& Account creation date & Distributions of account creation date.\\
 & Distribution of followers & Distributions of the numbers of followers.\\
 & Distribution of friends & Distributions of the numbers of friends.\\
 & Distribution of age and gender & Distribution of gender and age categories.\\

\hline
\end{tabular} 
\caption{\footnotesize List of evaluation criteria for comparing various methods of creating a national-sample of Twitter users from tweet and account aspects.}
\label{tab:evaluation} 
\end{table*} 
 
\begin{table*}[t]
\centering 
\footnotesize
\begin{tabular}{m{1.8cm}m{5.5cm}m{7.5cm}}
\hline \\[-1.8ex]
Category & Criteria & Description \\
\hline
  & MAPE where N $\sim$ M & Base model that uses only the total population count from the census (N) and Twitter (M).\\
  & MAPE where N $\sim \sum_g$ M(g) & Uses gender marginal counts only.\\
 Population-Level & MAPE where N $\sim \sum_a$ M(a) & Uses age marginal counts only.\\
  & MAPE where N $\sim \sum_{a,g}$ M(a, g) & Uses the joint distributions inferred from Twitter but only the total population counts from the census.\\
  & MAPE where log N(a, g) $\sim$ log M(a, g) + a + g & Uses the joint distributions inferred from Twitter and the joint histograms from the census.\\
\hline
\end{tabular} 
\caption{\footnotesize List of evaluation criteria for comparing various methods of creating a national-sample of Twitter users from population aspect.}
\label{tab:evaluation population} 
\end{table*} 
 
For the population-level metric, our objective is to determine which of the four Twitter sampling methods is most effective for generating a nationally representative sample of users. To achieve this, we adopt a test outlined in \cite{wang2019demographic} and employ the representative samples detailed in Section \nameref{sec:debiasing} to estimate the overall population of the United States based on Twitter data. In essence, we conduct a regression analysis that correlates the actual population sizes of various areas within the United States (such as states, divisions, or regions) with the number of American Twitter users from different age and gender groups in those specific locations. This analysis helps us assess the representativeness of the Twitter data for estimating the US population.

In a more detailed breakdown, we compare five distinct models that rely on different data sources and operate under different assumptions. The first model (N $\sim$ M) serves as the baseline and utilizes solely the total population count obtained from the census data along with Twitter user data, without applying any debiasing coefficients. The subsequent three models are grounded on the assumption of homogeneous inclusion probabilities:
The second model (N $\sim \sum_g$ M(g)) uses only gender-specific marginal counts.
The third model (N $\sim \sum_g$ M(a)) uses only age-specific marginal counts.
The fourth model (N $\sim \sum_{a,g}$ M(a, g)) employs the joint distribution inferred from Twitter data alongside the total population counts from the census data. Finally, the fifth model (log N(a, g) $\sim$ log M(a, g) + a + g) leverages the joint distribution inferred from both Twitter data and census data.
For each of these five prediction tasks, we assess their performance using the mean absolute percentage errors (MAPE) evaluation metric, calculated as specified in Equation \ref{equ:mape}. In this equation, $\hat{N}_i$ represents the predicted population size, $N_i$ is the actual population size, and the summation is performed over all geographical units of interest, such as states, regions, or divisions in the United States.

\begin{equation}
      MAPE(N) = \frac{100\%}{n} \sum_i^{Geo.} \frac{|\hat{N}_i - N_i|}{|N_i|}
\label{equ:mape} 
\end{equation}


\section{Results}
The following subsections provide a detailed presentation of our results by comparing the four Twitter sampling methods. First, we present essential statistics regarding the outcomes of each sampling method. Second, in accordance with the methodology outlined in Section \nameref{sec:prepro}, we randomly sample 30K accounts from the output of each sampling method. Subsequently, we apply pre-processing filters and select a random sample of 10K users from the remaining pool. We then report the metrics at both tweet- and user-levels for this subset. 
Lastly, we generate debiased samples from each 10K random sample by computing inclusion probabilities. We compare their mean absolute percentage errors (MAPE) for the task of estimating the United States population using Twitter data.

\subsection{Tweet-Level and User-Level Metrics}
In Table \ref{tab:tweet-level result}, we provide a comparison of various tweet-level metrics across the four Twitter sampling methods employed to create random user samples from a country. These metrics include the number of tweets, the count of unique users, the average number of tweets per account, and the percentage of tweets in English. The results show that the \textit{bounding box} (BB) and \textit{location query} (Loc) sampling methods produce a significantly higher number of tweets compared to the \textit{language query} (Lang) and \textit{1\% steam} methods. Among the four Twitter sampling methods, BB and Loc methods produce more than 18 million tweets, whereas the Lang and 1\% stream methods generate 4.5 million and 174,000 tweets, respectively, within the same timeframe. The same pattern is observed when examining the number of unique users and the average tweets per account metrics, except that the difference between BB and Loc methods and Lang method is notably smaller than their difference in the number of tweets. This suggests that the BB and Loc methods have a higher rate of account duplication compared to the other two sampling methods. Lastly, among the three methods that do not explicitly filter for language, the BB method has a higher proportion of English tweets.

\begin{table*}[t]
\centering
\begin{tabular}{m{5cm}m{2cm}m{2cm}m{2cm}m{2cm}}
\hline
 & \textit{BB} & \textit{Loc} & \textit{Lang} & \textit{1\%} \\ \hline
Number of tweets & 18,181,424 & 18,804,550 & 4,508,702 & 174,084 \\
Number of accounts & 728,028 & 738,595 & 425,041 & 94,250 \\
Average tweets per account & 24.974 & 25.46 & 10.608 & 1.847 \\
Ratio of tweets in English & 0.823 & 0.808 & 1 & 0.807 \\
\hline
\end{tabular}
\caption{\footnotesize Statistics of the number of tweets and users collected by each sampling method. }
\label{tab:tweet-level result} 
\end{table*}

Due to the computational expense of bot detection and age, gender, and location inferences, our aim is to compare a random sample of 10K users from each Twitter sampling method. To ensure that we have at least 10K users from each method after applying pre-processing steps, we initially select a random sample of 30K users from each Twitter sampling method. We then execute all pre-processing steps on this larger sample and subsequently select a random sample of 10K users from the remaining pool. The table in Table \ref{tab:filter_changes} provides information about the number of accounts that have been removed after each filter has been applied to the 30K-sample.

\begin{table*}[!h]
\centering
\begin{tabular}{m{7cm}m{1cm}m{1cm}m{1cm}m{1cm}m{1cm}}
        \hline \\[-1.8ex] 
        \multirow{2}{*}{Filter method}  &\multicolumn{5}{c}{Number of accounts removed due to each filter}   \\
        \cmidrule(r){2-6}
         & BB & Loc & Lang & 1\% \\
        \hline \\[-1.8ex] 
    Verified accounts &	792 &	705 &	744 & 862 \\
    Accounts with less than 100 tweets &	3,971 &	3,991 &	2,650 & 1,268 \\
    Accounts with less than 9 month age &	830 &	764 &	928 & 1,073 \\
    Accounts has keywords in Description &	793 &	675 &	719 & 768 \\
    Non-English Accounts &	4,610 &	4,433 &	0 & 5,004 \\
    Non-US Accounts Accounts &	92 &	0 &	0 & 0 \\
        \hline \\[-1.8ex]

\end{tabular}
\caption{\footnotesize Number of Accounts Removed by Each Filtering Method in USA 30K Samples }
\label{tab:filter_changes}
\end{table*}

\begin{figure}[!h]
    \centering
    \begin{subfigure}{0.2\textwidth}
        \caption{}
        \label{fig:tweet_dist}
        \includegraphics[scale=0.25]{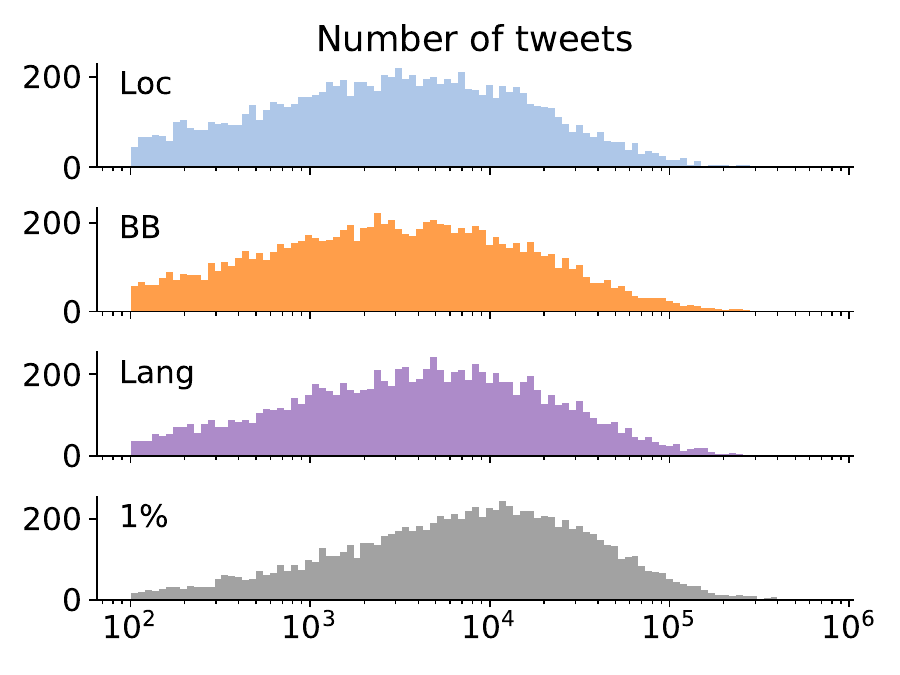}
    \end{subfigure}
    \begin{subfigure}{0.2\textwidth}
        \caption{}
        \label{fig:tweet_per_day}
        \includegraphics[scale=0.25]{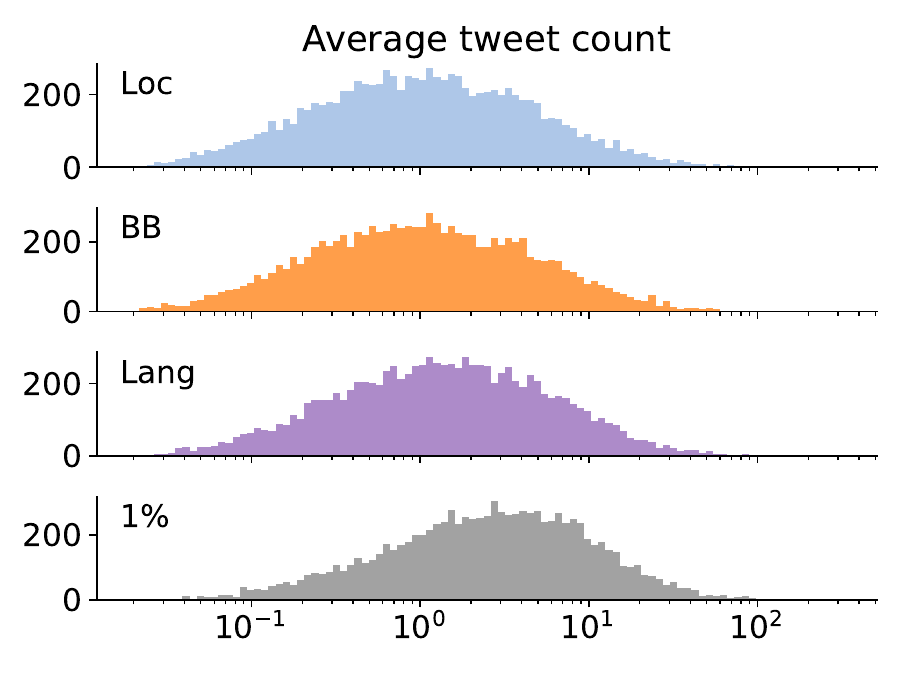}
    \end{subfigure}
    \\
    \begin{subfigure}{0.2\textwidth}
        \caption{}
        \label{fig:like}
        \includegraphics[scale=0.25]{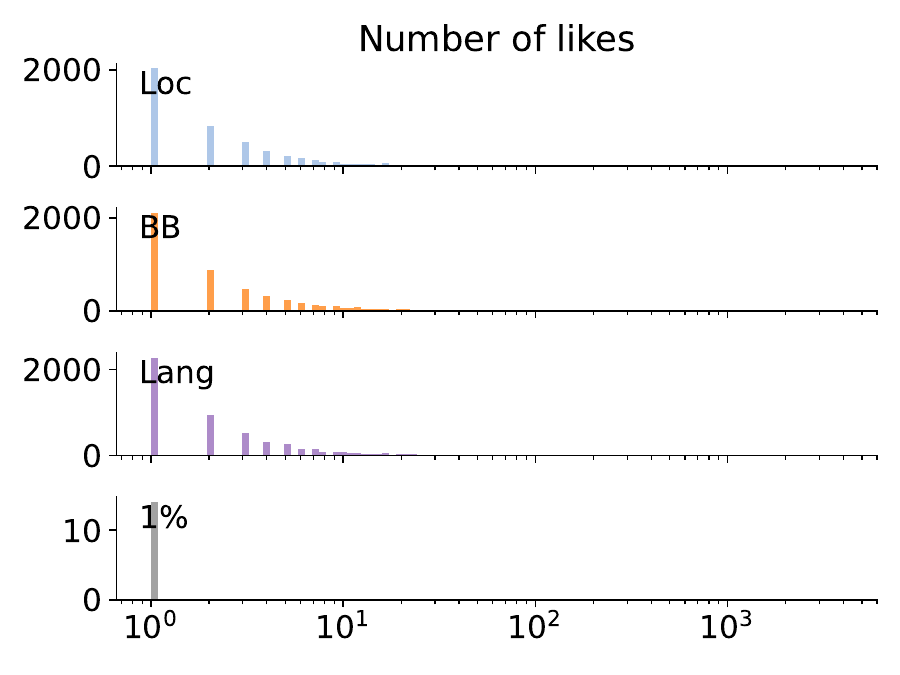}
    \end{subfigure}
    \begin{subfigure}{0.2\textwidth}
    \caption{}
        \label{fig:tenure}
        \includegraphics[scale=0.25]{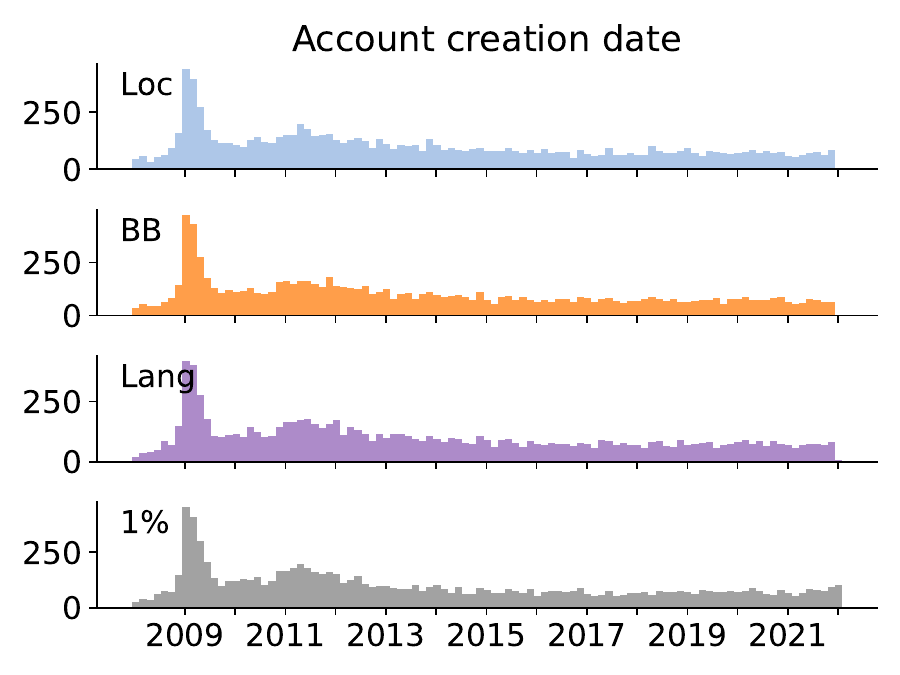}
    \end{subfigure}
    \\
    \begin{subfigure}{0.2\textwidth}
        \caption{}
        \label{fig:follower}
        \includegraphics[scale=0.25]{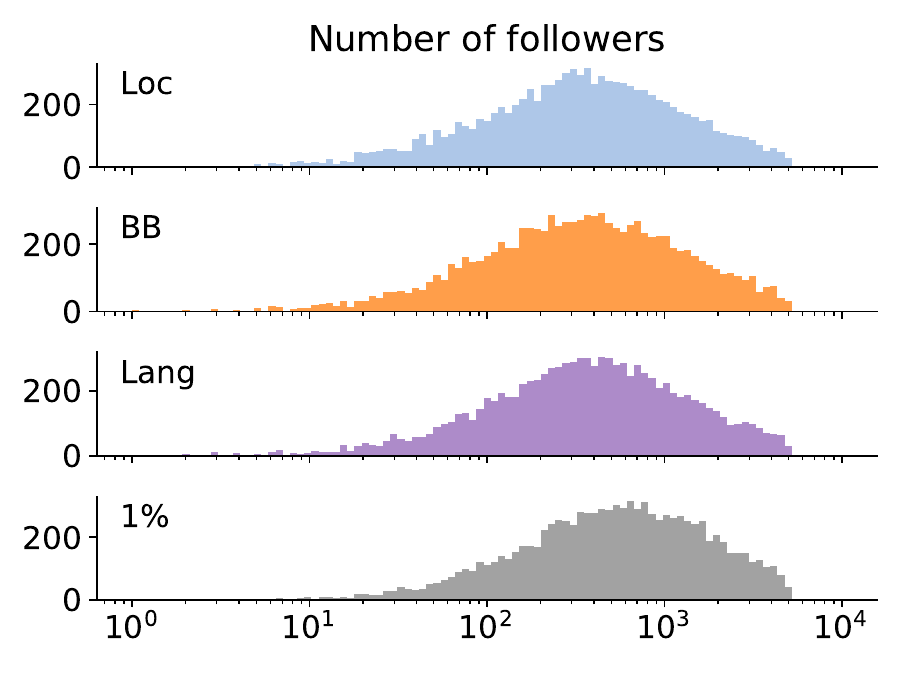}
    \end{subfigure}
    \begin{subfigure}{0.2\textwidth}
        \caption{}
        \label{fig:following}
        \includegraphics[scale=0.25]{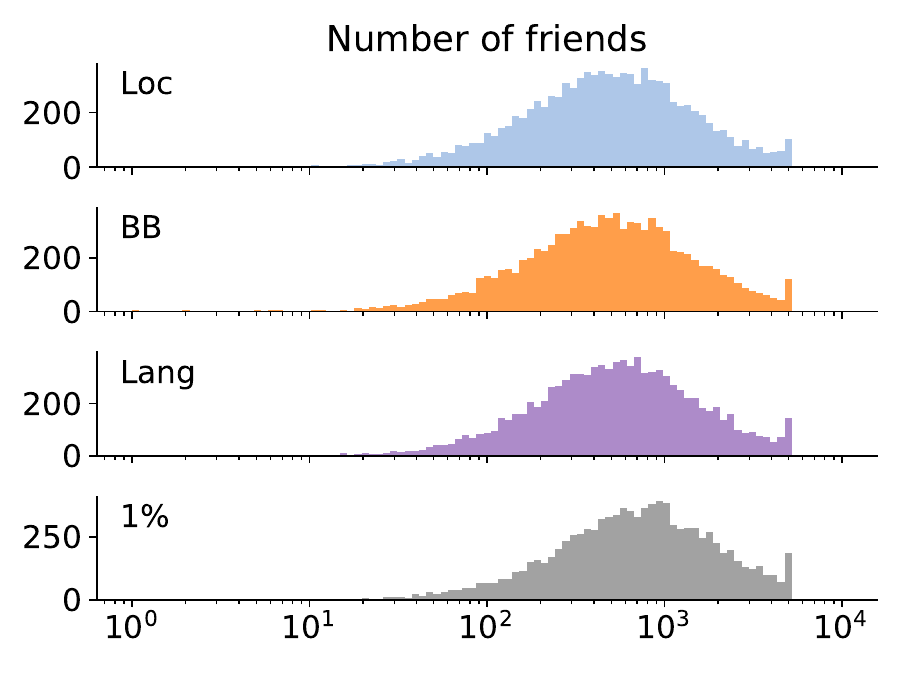}
    \end{subfigure}
    \\
    \begin{subfigure}{0.2\textwidth}
        \caption{}
        \label{fig:gender}
        \includegraphics[scale=0.25]{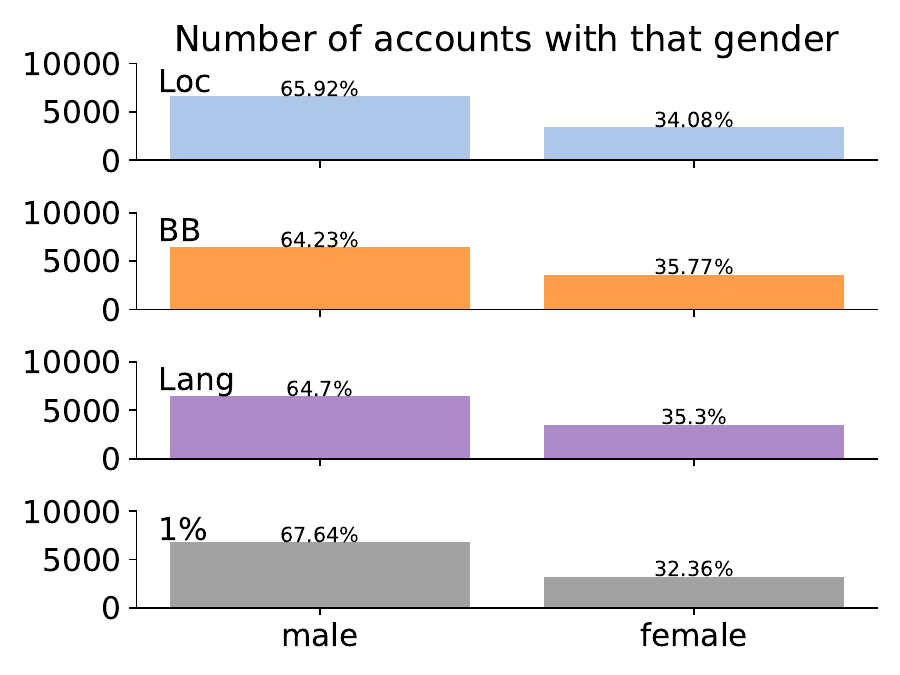}
    \end{subfigure}
    \begin{subfigure}{0.2\textwidth}
        \caption{}
        \label{fig:age}
        \includegraphics[scale=0.29]{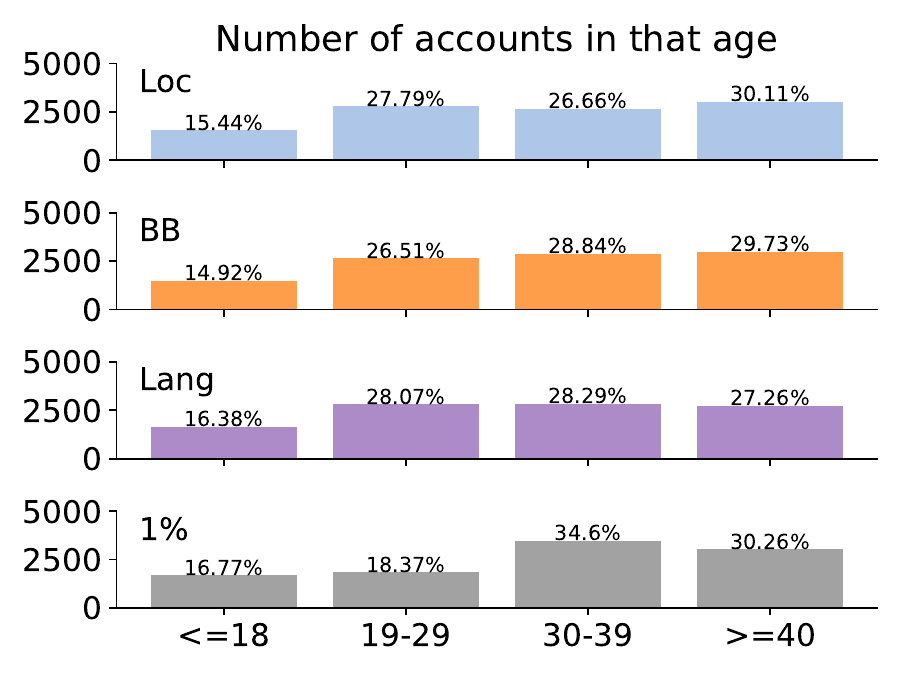}
    \end{subfigure} 
\caption{\footnotesize Distributions of (A) number of tweets; (B) average number of tweets per day; (C) number of likes; (D) account creation date; (E) number of followers; and (F) number of friends for different groups. Distribution of users with respect to (G) gender and (H) age across the four Twitter sampling methods.}
\label{fig:accountcompare}
\end{figure}

\begin{figure}[h]
    \centering
    \begin{subfigure}{0.2\textwidth}
        \caption{}
        \label{fig:tweet_dist_P-values}
        \includegraphics[scale=0.3]{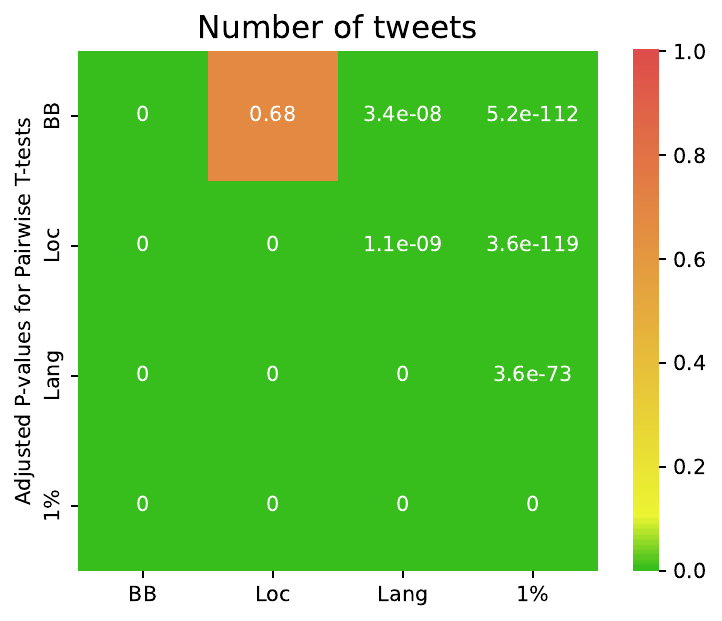}
    \end{subfigure}
    \begin{subfigure}{0.2\textwidth}
        \caption{}
        \label{fig:tweet_per_day_P-values}
        \includegraphics[scale=0.3]{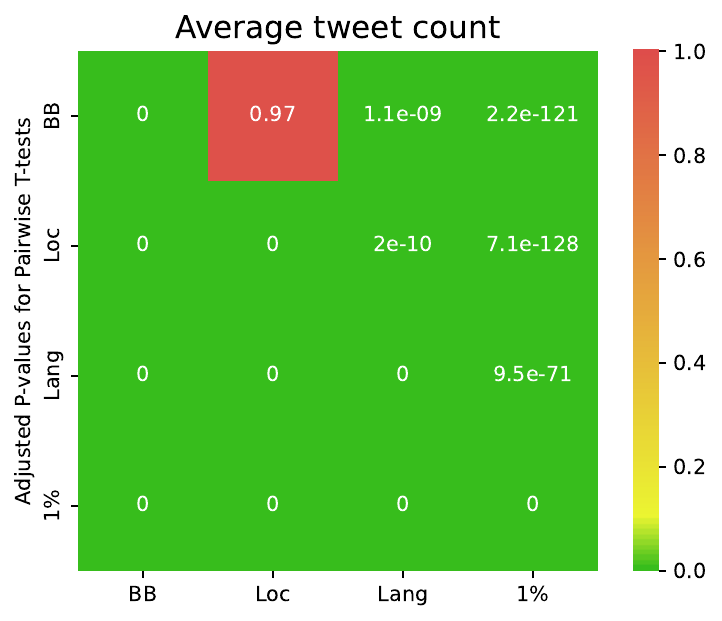}
    \end{subfigure}
    \\
    \begin{subfigure}{0.2\textwidth}
        \caption{}
        \label{fig:like_P-values}
        \includegraphics[scale=0.3]{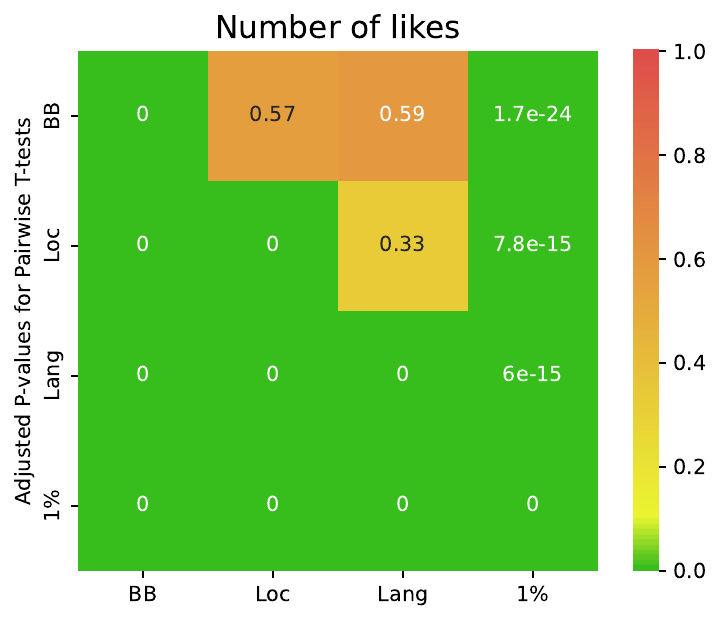}
    \end{subfigure}
    \begin{subfigure}{0.2\textwidth}
    \caption{}
        \label{fig:tenure_P-values}
        \includegraphics[scale=0.3]{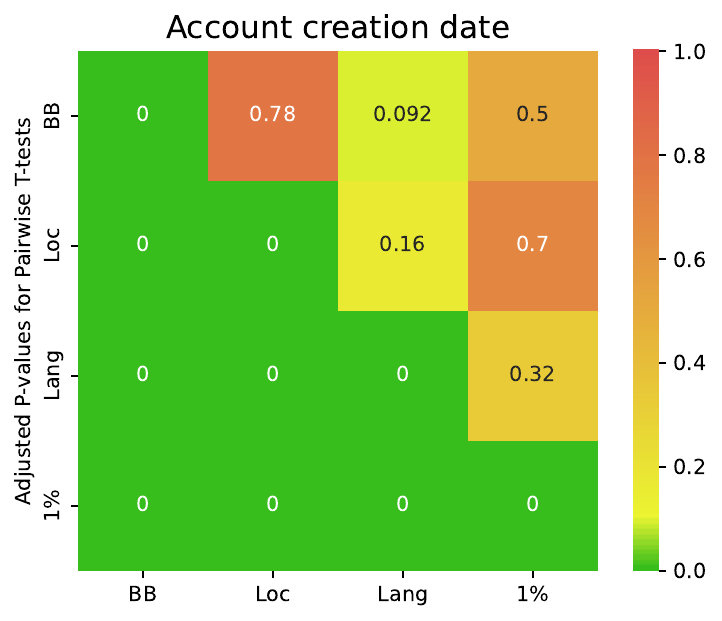}
    \end{subfigure}
    \\
    \begin{subfigure}{0.2\textwidth}
        \caption{}
        \label{fig:follower_P-values}
        \includegraphics[scale=0.3]{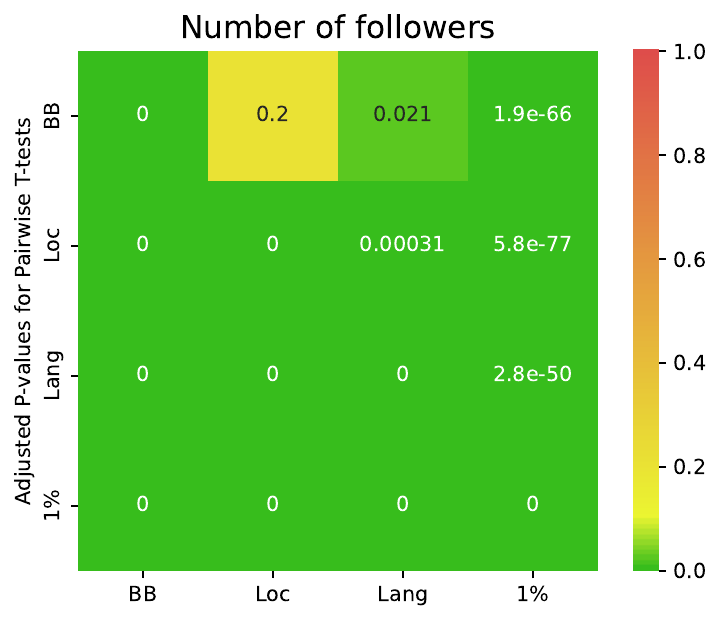}
    \end{subfigure}
    \begin{subfigure}{0.2\textwidth}
        \caption{}
        \label{fig:following_P-values}
        \includegraphics[scale=0.3]{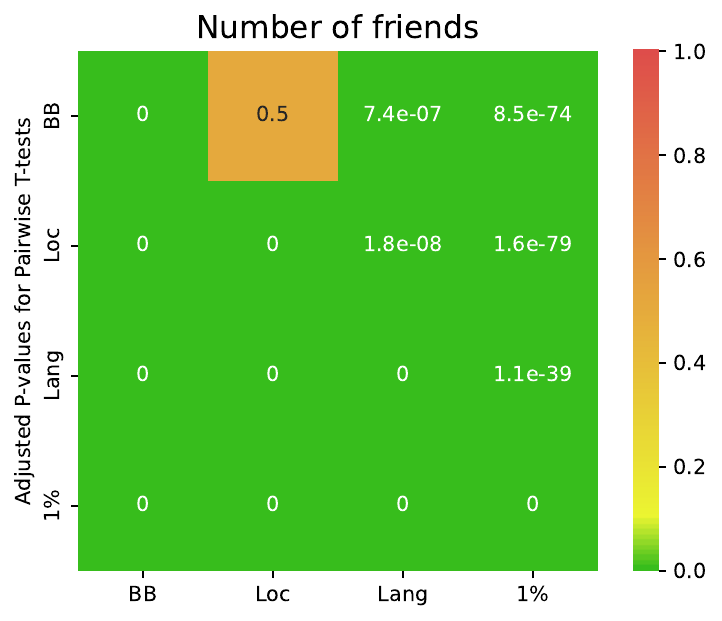}
    \end{subfigure}
    
\caption{\footnotesize Heatmap of P-values for Pairwise T-tests of  (A) number of tweets; (B) average number of tweets per day; (C) number of likes; (D) account creation date; (E) number of followers; and (F) number of friends across the four Twitter sampling methods.}
\label{fig:accountcompare_P-values}
\end{figure}

The comparison results of tweet- and user-level metrics are reported in Fig \ref{fig:accountcompare} and Tables \ref{tab:tweet_level_metrics} and \ref{tab:user_level_metrics}, and the corresponding $t$-test results are illustrated in Fig \ref{fig:accountcompare_P-values}. More specifically, the distributions of the total number of tweets are presented in Fig \ref{fig:tweet_dist}. Notably, the \textit{1\% stream} method generates more tweets ($M$ = 19,873.9, $p$ = 0.00) compared to the other methods (see Table \ref{tab:tweet_level_metrics} and Fig \ref{fig:accountcompare_P-values}). Additionally, since the number of generated tweets depends on account age, we also depict the distributions of the number of tweets per day in Fig \ref{fig:tweet_per_day}. This metric is calculated by dividing a user's total number of tweets by the number of days since their account creation. Once again, we observe that users from the \textit{1\% stream} method tend to tweet more frequently (M = 5.81, $p$ = 0.00) than those from other methods (see Table \ref{tab:tweet_level_metrics}). We have also conducted a comparison of the number of likes across the four Twitter sampling methods and have depicted the corresponding distributions in Fig \ref{fig:like}. It's worth noting that the \textit{BB}, \textit{Loc}, and \textit{Lang} methods exhibit nearly identical distributions. However, users sampled from the \textit{1\% stream} method tend to have significantly fewer likes (M = 0.00, $p$ = 0.00). This discrepancy is primarily attributed to the fact that the \textit{1\% stream} method collects data in real-time, often when engagements with posts have just commenced. This is in contrast to the other methods, which may include tweets posted up to the past seven days, allowing for more time for engagements to accumulate.
Additionally, we present the distributions of account creation times in Fig \ref{fig:tenure}. Across all methods, we notice a peak around the year 2009, likely reflecting the rapid growth of Twitter during that period \cite{yang2022twitter}. There is also an increase in the number of created accounts in 2011, coinciding with a period of significant user growth on Twitter\footnote{\url{https://www.theguardian.com/technology/pda/2011/sep/08/twitter-active-users}}. In comparison to other methods, the \textit{1\% stream} method appears to generate more younger accounts ($p$ = 0.00), specially those that were created in 2022. However, for the remaining time period, the distributions appear similar across all four methods.

Figures \ref{fig:follower} and \ref{fig:following} display the distributions of the numbers of followers and friends, respectively. These figures illustrate that users in the \textit{1\% stream} method tend to have slightly more followers (M = 911.3, $p$ = 0.00) and friends (M = 1059.6, $p$ = 0.00) compared to users in the other sampling methods (see Table \ref{tab:user_level_metrics} and Fig \ref{fig:accountcompare_P-values}). Interestingly, in Fig \ref{fig:following}, it becomes evident that nearly half of the users in the \textit{1\% stream} sample have over 1,000 friends. Moreover, the \textit{1\% method} exhibits a higher number of accounts with approximately 5,000 friends or more. It's worth noting that the peak around 5,000 friends in Fig \ref{fig:following} is attributed to a Twitter anti-abuse limitation, which stipulates that an account cannot follow more than 5,000 friends unless it has more than 5,000 followers\footnote{\url{https://help.twitter.com/en/using-twitter/twitter-follow-limit}}. This policy leads to the observed distribution pattern. 

As outlined in Section \nameref{sec:debiasing}, age, gender, and location are critical demographics for constructing a nationally-representative sample of individuals. Consequently, it is essential to compare the user samples generated by each sampling method in terms of these three metrics. Fig \ref{fig:accountcompare} displays the distributions of age and gender across the four Twitter sampling methods. When examining gender distributions (Fig \ref{fig:gender}), two notable observations emerge: (1) All four methods yield unbalanced gender data, with a majority of users (over 64\%) being male, (2) The \textit{1\% stream} method produces slightly fewer women compared to the other three methods, which generate a nearly equal fraction of women. 

Regarding the age (Fig \ref{fig:age}), the percentage of users who are under 18 or over 40 is nearly identical across all methods. However, there is a significant difference for the 19-29 and 30-39 age cohorts in the \textit{1\% stream} method. While the other three methods generate approximately 27\% of users in both the 19-29 and 30-39 age groups, the \textit{1\% stream} method has only 18\% of users in the 19-29 cohort and 35\% in the 30-39 cohort, indicating a notable deviation.

\begin{table*}[!h]
    \centering
    
    \caption{\footnotesize Average and standard deviation of the number of tweets and likes across the four Twitter sampling methods}
    \begin{tabular}{m{2cm}m{1.2cm}m{2cm}m{2.2cm}m{2 cm}m{1.1cm}m{2cm}}
        \hline \\[-1.8ex] 
        \multirow{2}{*}{}  &\multicolumn{2}{l}{Number of tweets }&\multicolumn{2}{l}{ Average tweet count }&\multicolumn{2}{l}{Number of likes }   \\
        \cmidrule(r){2-7}
         & Mean & Std  & Mean & Std  & Mean & Std  \\
        \hline \\[-1.8ex] 
    BB & 10,370.6 &	23,388.4 &	2.97 & 6.86 & 3.73 & 36.44  \\
    Loc &	10,235.9 &	22,028.5 &	2.96 & 6.31 & 3.40 & 43.70 \\
    Lang &	12,210.7 &	23,731.2 &	3.59 & 7.53 &  4.07 & 52.05  \\
    1\%Stream  &	19,873.9 &	34,870.5 &	5.81 & 9.92 &  0.00 & 0.04  \\

        \hline \\[-1.8ex]
    \end{tabular}
    \label{tab:tweet_level_metrics} 
\end{table*}

\begin{table}[h]
    \centering
    
    \caption{\footnotesize Average and standard deviation of the number of followers and friends across the four Twitter sampling methods.}
    \begin{tabular}{m{1.7cm}m{1cm}m{1.5cm}m{1cm}m{1cm}}
        \hline \\[-1.8ex] 
        \multirow{2}{*}{}  &\multicolumn{4}{c}{Number of followers\> Number of friends }   \\
        \cmidrule(r){2-5}
         & Mean & Std  & Mean & Std   \\
        \hline \\[-1.8ex] 
    BB &  683.6 & 869.3 & 804.2 & 913.9 \\
    Loc  & 668.1 & 849.7 & 795.6 & 902.4 \\
    Lang  & 712.4 & 888.7 & 870.1 & 966.2 \\
    1\% Stream  & 911.3 & 989.9 & 1059.6 & 1060.2 \\
    
        \hline \\[-1.8ex]
    \end{tabular}
    \label{tab:user_level_metrics}  
\end{table}



\begin{figure}[H]
    \centering
    \includegraphics[scale=0.07]{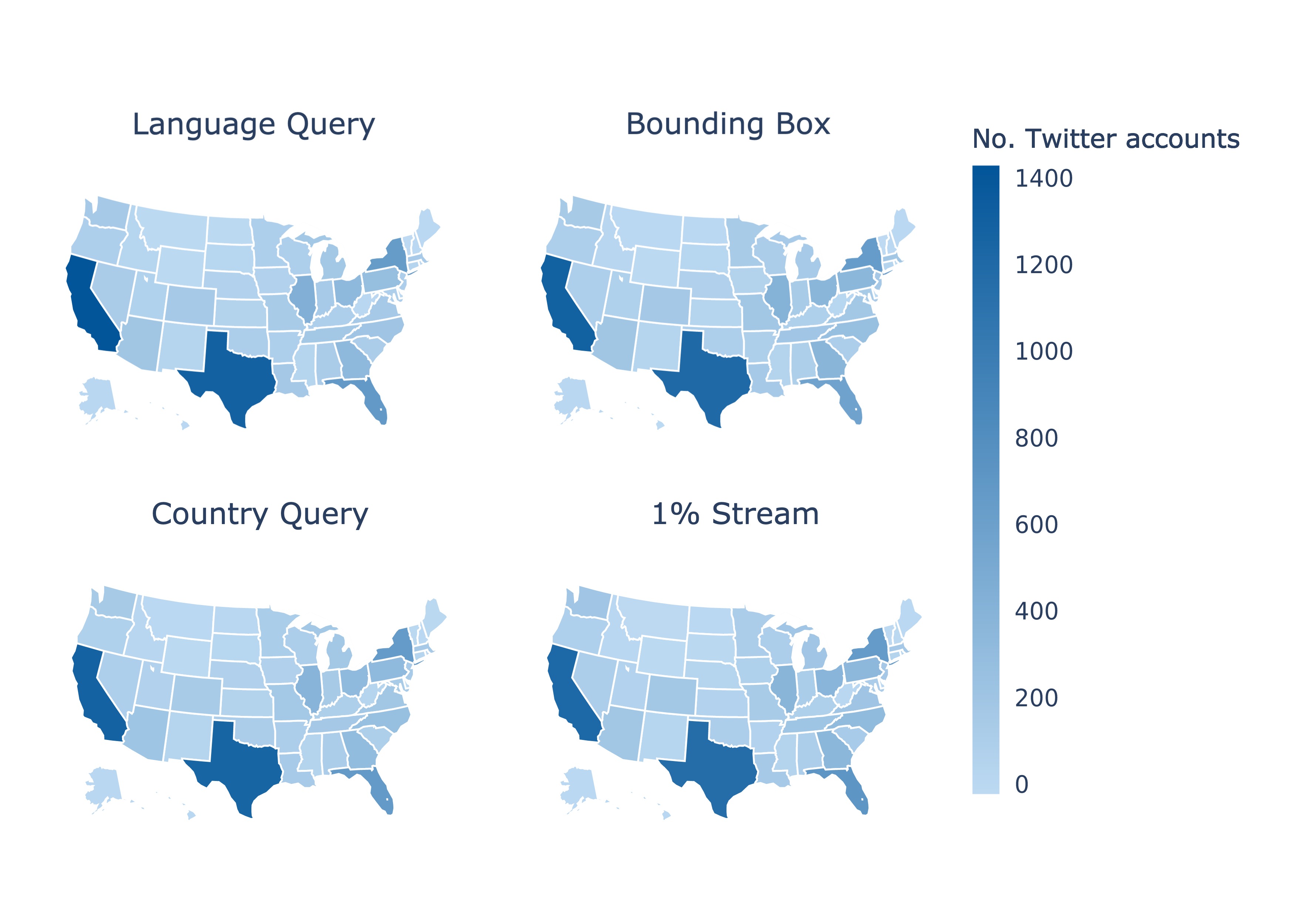}
    \caption{\footnotesize Map of the number of users located in US states. All four sampling methods produced at least one user in all 50 US states.}
    \label{fig:us_map}
\end{figure}

Finally, in Fig \ref{fig:us_map}, we present the distribution of the number of users located in each state within the United States. As discussed in Section \nameref{sec:debiasing}, we adopted the methodology proposed in \cite{barbera2019leads} to estimate the location of Twitter users at the US state level. In general, the distribution pattern is highly similar across all four Twitter sampling methods, with no significant variations between them. As expected, states with larger populations such as California, New York, Texas, and Florida have a higher number of Twitter users. Additionally, within our 10K-user sample, all four sampling methods managed to provide at least one user from each state, ensuring a diverse geographical representation.


\subsection{Population-Level Metrics}

The objective of this section is to assess the accuracy of the four debiased samples created from the four Twitter sampling methods in predicting the population of the United States using Twitter data. This prediction task is conducted at the state level, which necessitates having a sufficient number of users for all age and gender groups in all 50 US states. However, despite all four sampling methods providing at least one user in each state (as shown in Fig \ref{fig:us_map}), none of them yielded enough data to cover all combinations of age and gender across all US states. Table \ref{tab:state_missing} provides information on the number of US states where there are insufficient users to represent all age and gender groups, across different sample sizes (i.e., 5K, 8K, or 10K users) and Twitter sampling methods. As indicated in Table \ref{tab:state_missing}, even with a sample size of 10K users, there are still between 7 to 11 states lacking at least one demographic group (e.g., women aged 30 to 39 in the State of New York).

\begin{table}[t]
\begin{tabular}{@{}llll@{}}
\toprule
 Sampling Method & 5K Sample & 8K Sample & 10K Sample \\ \midrule
         1\% Stream & 14 & 12 & 10\\
         BB & 13 & 8 & 8\\
         Loc & 15 & 8 & 7\\
         Lang & 17 & 12 & 11\\
         \bottomrule
\end{tabular}
\caption{\footnotesize Number of US states that Twitter sampling methods did not generate enough users in all demographics groups.}
\label{tab:state_missing}
\end{table}

In our regression model for estimating the US population from Twitter data, each row corresponds to a specific demographic group within a particular state (e.g., men above 40 in New Jersey). If any demographic group is absent in the Twitter data for a specific state, all values in the corresponding row of the regression model would be zeros. However, it's important to note that no state has all demographic groups missing. Consequently, we made the decision to remove rows with all-zero values from our regression model. This led to the removal of 16 rows (3.92\%) from the \textit{bounding-box} sample, 12 rows (2.94\%) from the \textit{Country Query} sample, 16 rows (3.92\%) from the \textit{Language Query} sample, and 21 rows (5.15\%) from the \textit{1\% Stream} sample. It's worth mentioning that all of these samples consist of an equal size of 10K users each.

Following the approach outlined in \cite{wang2019demographic}, we assess the accuracy of the four representative samples using a leave-one-state-out cross-validation framework. In this evaluation, we calculate the mean absolute percentage error (MAPE) for the population estimates of the state that is left out of the analysis. The MAPE is computed using the formula specified in Equation \ref{equ:mape}. As mentioned in Table \ref{fig:accountcompare}, we calculate MAPE under five different scenarios. These scenarios encompass a baseline model that relies solely on the total population without the use of debiasing coefficients. Additionally, there are three models based on homogeneity, which consider whether to include both age and gender or only one of these variables. Lastly, there is a full model that takes into account heterogeneity by including both age and gender as factors in the analysis. 

\begin{figure*}[t]
    \centering
    \footnotesize
    \includegraphics[scale=0.28]{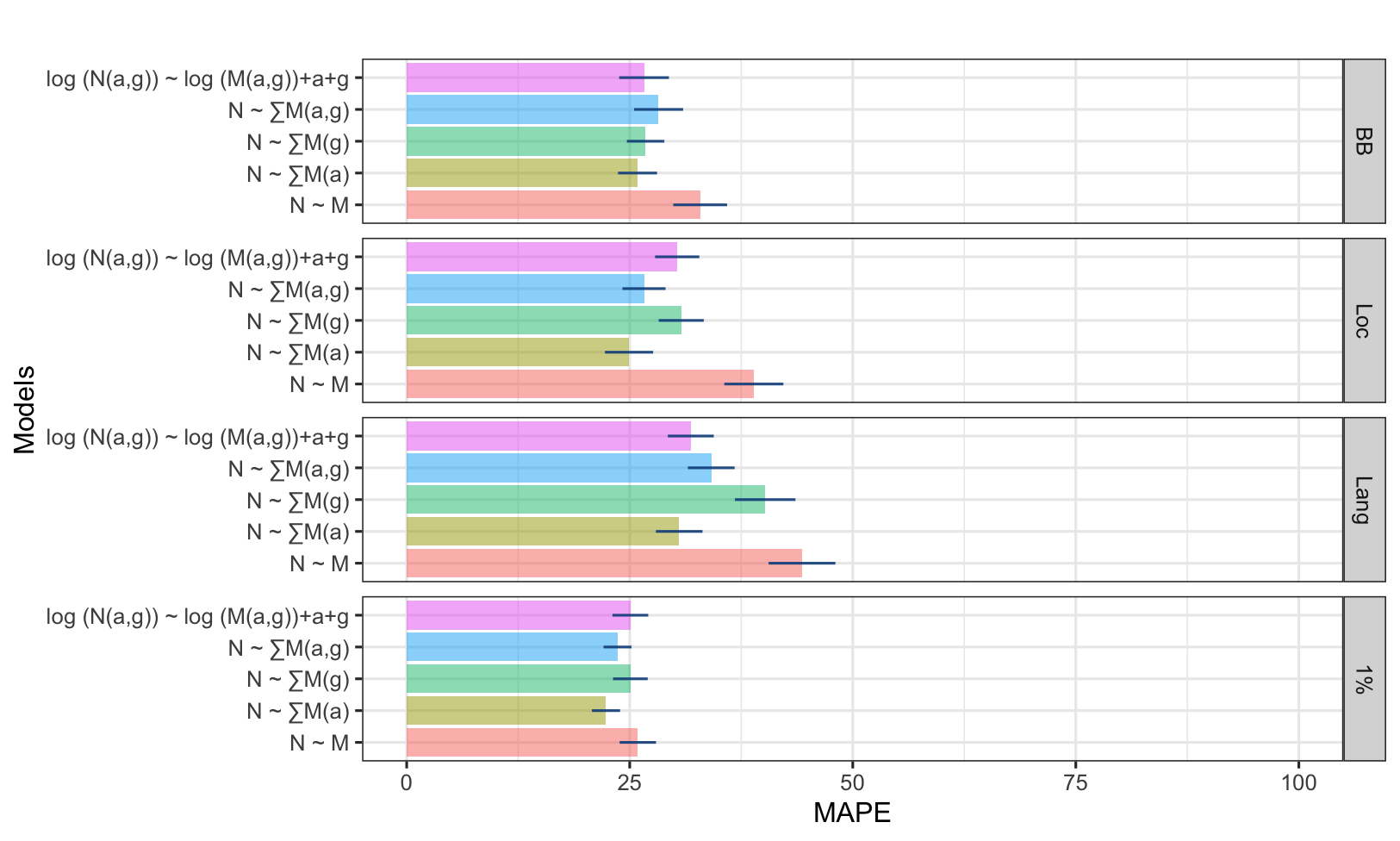}
    \caption{\footnotesize Performance on leave one state out population inference across different debiasing models where rows with all zero value were removed from the regression. The bar shows MAPE($N$) robust standard errors clustered on states.}
    \label{fig:mape}
\end{figure*}

In Figure \ref{fig:mape}, we observe that the results of the leave-one-state-out evaluation show a benefit to using the \textit{1\% Stream} Twitter sampling method. Across all five debiasing models, the \textit{1\% Stream} seems to achieve the minimum and the \textit{Language Query} method achieves the maximum prediction error. For the N $\sim$ M model, which is a baseline that does not use any debiasing, the \textit{1\% Stream} method achieves MAPE of 27\%, which is the least compared to \textit{BB}, \textit{Loc}, and \textit{Lang}, which achieved MAPE of 33\%, 39\%, and 45\%, respectively. The inclusion of the inferred age in the debiasing models N $\sim \sum$ M (a) decreases MAPE of the \textit{1\% Stream} method to 21\%, which again is the minimum among the \textit{BB} (26\%), \textit{Loc} (25\%), and \textit{Lang} (31\%) methods. The inclusion of the inferred gender in the debiasing models N $\sim \sum$ M (g) decreases the MAPE of the \textit{1\% Stream} method, but not as big as the inclusion of the age. Nonetheless, the \textit{1\% Stream} sample shows the minimum prediction error with MAPE of 25\% (the \textit{BB}, \textit{Loc}, and \textit{Lang} methods obtained MAPE of 27\%, 30\%, and 41\% respectively). Same pattern holds true for N $\sim \sum$ M (a, g) and logN(a,g) $\sim$ logM(a,g) + a + g models. Moreover, the results show that even the baseline model of the \textit{1\% Stream} sample outperforms the other three sampling methods in all five modelling scenarios.  

\subsection{Robustness Tests}

To ensure that the observed results are not the artifact of our regression settings, in which we removed the rows with all zero values, or artifact of our pre-processing decisions, we replicate the results of Fig \ref{fig:mape} for the following scenarios: 1) including the District of Columbia to the 50 states and remove rows with zero values in the regression model (see Fig \ref{fig:mape-51} in Appendix), 2) removing all states that their Twitter data are missing at least one demographic group from the Twitter and census data and measure MAPE for the remaining states (see Fig \ref{fig:mape-39} in Appendix), 3) aggregate data at the nine US divisions (the U.S. Census Bureau groups the 50 states and the District of Columbia into four geographic regions and nine divisions based on geographic proximity\footnote{\url{https://www.census.gov/programs-surveys/economic-census/guidance-geographies/levels.html}}), which results in no missing demographic group at all nine divisions (see Fig \ref{fig:mape-div} in Appendix), 4) removing users with less than 200 tweets in the pre-processing step instead of the original 100 tweet threshold (note that due the recent restrictions imposed on Twitter API, we are not able to test the lower threshold because it will produce new users, for which, we cannot obtain bot score, age, gender, and location) (see Figures \ref{fig:accountcompare200} and \ref{fig:mape200} in Appendix), and 5) removing users with the age of less than one year instead of the 9 month threshold (see Figures \ref{fig:accountcompare12} and \ref{fig:mape12} in Appendix). 

Across all of these five robustness tests, the overall qualitative patterns of the comparison between the four Twitter sampling methods remain the same, with the \textit{1\% Stream} sampling method obtaining the minimum prediction error on the US population inference task. Interestingly, for the leave-one-division-out cross-validation setting, we see that the error rates drop nearly in half across the four sampling methods and five population inference tasks (Fig \ref{fig:mape-div}). This error reduction suggests that knowledge of the nine US division specific platform biases is more important than the state-specific ones for accurate estimates of the country population. In other words, to achieve a good performance, a model at least need to be exposed to some divisions within the US during its training period.




\section{Discussion}

Twitter has become the most studied social media platform where people express their everyday opinions, especially about politics. This fact encouraged a wealth of social and computer science scholars to use Twitter data for measuring national-level statistics for political outcomes, health metrics, or public opinion research. However, Twitter is not a representative sample of the population due to demographic imbalance in usage and penetration rates. To address this fact, researchers often try to create a random sample of Twitter users from a country. However, at least four widely-used sampling methods exist in the literature, and the extent to which their outputs are similar or different has not been explored systematically so far.

In this paper, we tackled this issue by comparing the performance of the four different Twitter sampling methods on some carefully devised evaluation metrics. More specifically, the four methods include (1) \textit{1\% stream}, in which one uses Twitter Stream API to get 1\% of tweets in real-time and then sample from the authors of the tweets, (2) \textit{location query}, in which one uses Twitter Search API and query for a country of interest, (3) \textit{language query}, in which one uses Twitter Search API, query for language(s) representing the country of interest, and filter for the country, and (4) \textit{bounding-box}, in which one uses the `bounding-box' field in the Search API and query for the coordinates enclosing the country of interest. After carefully reviewing the literature, we devised three tweet-level, eight user-level, and five population-level evaluation metrics to compare these four Twitter sampling methods.

Our results highlight the \textit{1\% Stream} Twitter sampling method, which exhibits different characteristics compared to the other three sampling methods. To explain the implications of differences of  \textit{1\% Stream} sample and other samples, The \textit{1\% Stream} method involves taking a 1\% sample of tweets and then collecting a 10,000 user sample from them. On the other hand, the Lang method takes a 10,000 user sample from all users. While the filters applied in both methods are the same, the \textit{1\% Stream} method may result in a sample that is skewed towards more active users. This is because the more a user tweets, the more likely they are to be included in the 1\% sample, from which the 10,000 user sample is taken. In contrast, other three methods(e.g. the \textit{Lang} method) provide an equal probability for all users to be included in the sample. More particularly, Twitter users collected by the \textit{1\% Stream} method tend to have more tweets, tweets per day, followers, and friends, and fewer number of likes. In addition, it appears that the \textit{1\% Stream} sampling method provides slightly younger accounts (i.e. accounts created around 2022), slightly more male users, significantly fewer users in the 19-29 age stratum, and significantly more users in the 30-39 age stratum, compared to the other three sampling methods.

More importantly, the \textit{1\% Stream} method achieves the minimum error, compared to the other three methods, in the prediction task of estimating the population of the US from Twitter users. This is true across five different debiasing models, each attempting to make the sample representative of the US population. A baseline model using the \textit{1\% Stream}, in which we do not implement any debiasing technique, outperforms or equates all debiased forms of the other sampling methods in terms of the prediction error (except for the \textit{Location Query} method when using marginal age counts).

However, the \textit{1\% Stream} Twitter sampling method has with some disadvantageous. First, it is time-consuming because the Twitter Stream API provides tweets in real-time. This means that, for example, if one needs to collect a month of stream data, she cannot get it immediately and has to wait for a whole month. This is not true for the other sampling methods, in which one can get the same period of data in a few hours or days, depending on the size of the country. Second, and closely related to the first disadvantage, one cannot obtain the stream data for past time periods unless she had started to do so before. Finally, the \textit{1\% Stream} method is unsuitable for studies focusing on user engagement metrics (i.e. number of likes, retweets, comments, and views). This is due to the fact that this method collects tweets in real-time, which in most of the cases, as we have shown in Fig \ref{fig:like}, it ends up catching very recently posted tweets, which in turn means they have not yet been seen by most of the other users, and therefore not acquired user engagements.

The drawbacks of the \textit{1\% Stream} Twitter sampling method underscore the importance of the second-best sampling method in our study, which is the \textit{bounding box} method. While its results are identical to \textit{Location Query} and \textit{Language Query} methods in terms of tweet- and user-level metrics (see Figs \ref{fig:accountcompare}), it clearly outperforms them with respect to the prediction error.

While Twitter has been one of the most studied social media platforms, the advantages and disadvantages of different sampling strategies remain unclear. Our results illuminate the positive and negative characteristics of the four main sampling methods used in the literature and help researchers choose the one that best suits their research goals and designs. By identifying the best sampling methods, our results also pave the way for conducting more accurate social listening studies and building more accurate machine learning models. Moreover, our approach and results may be adapted to conduct similar studies for other social media platforms.

The performance of the \textit{1\% Stream} and \textit{bounding box} Twitter sampling methods is better compared to the other two sampling methods when used for estimation of the population of US. Future research should explore the role of large language models in improving the performance of or automating usage of different models used in the process of computing the inclusion probabilities \cite{cerina2023artificially}, determine significant parameters in the calculation of inclusion probabilities \cite{alizadeh2011determining}, and test the temporal and regional validity of the results.

\section{Acknowledgment}
This project received funding from the European Research Council under the EU’s Horizon 2020 research and innovation program (Grant Agreement No. 883121). We thank conference participants at the IC2S2 2023 and APSA 2023 conferences for their helpful feedback. We thank Sara Yari Mehmandoust, Mahdis Abbasi, Sima Mojtahedi, Mohammad Hormati, and Zahra Baghshahi for outstanding research assistance.

\section{Conflict of Interest}
The authors declare no conflict of interest.

\section{Data Availability Statement}
The datasets generated and analyzed during the current study are not publicly available due to the Twitter’s Developer Agreement. However, the user IDs and tweet IDs are available from the corresponding author on request.

\bibliography{references}

\appendix

\pagenumbering{arabic}
\renewcommand{\thesection}{S\arabic{section}}
\setcounter{figure}{0} \renewcommand{\thefigure}{S\arabic{figure}}
\setcounter{table}{0} \renewcommand{\thetable}{S\arabic{table}}

\section{Random User ID Generation Sampling Method}
\label{sec:appendix_data_collection}
 Twitter has two distinct types of IDs - the older 32-bit serial IDs and the newer 64-bit random IDs. To ensure the validity of the sample, we developed a function that is capable of generating both types of IDs. 
 
 To gather a representative sample of user IDs, we use an iteration function that generates both old 32-bit IDs and new 64-bit IDs. For each iteration, we set a random seed generated from a seed generator. This ensures reproducibility if the process is rerun on any system. During each iteration, we generate 1000 IDs, with 5 being old IDs and 995 being new IDs. The distribution is heavily skewed towards new IDs due to their higher hit rate compared to old IDs. The old IDs are simply random 32-bit integers that have not been generated before. For 64-bit IDs, we employ the snowflake approach, which generates unique IDs and is also used by Twitter \footnote{https://developer.twitter.com/ja/docs/basics/twitter-ids}. This approach generates valid user IDs while avoiding reserved IDs for future use. Twitter created the Twitter Snowflake ID generator to produce unique IDs for objects needing identification, such as tweets and users. Twitter Snowflake optimizes the traditional UUID concept to work on a large scale while maintaining sortable IDs and manageable sizes. Twitter Snowflake IDs are 64-bit long and sortable by time, making them ideal for indexing and efficient in storage and processing. The system is scalable, accommodating up to 1024 machines, and highly available, with each machine capable of generating 4096 unique IDs per millisecond. The fact that the IDs are time-dependent makes it possible to generate working IDs with a higher chance since one can restrict the generation to only IDs from the past \cite{Snowflake_ID_Generator}. 
 
 We Found 357K users with Random ID and filtered the American users with query \textit{place country:US}. However, we got only 127 users from the US. Finding working IDs from a specific country with a country code is tricky. Thus, getting a large enough sample from random user IDs is almost impossible due to the low number of calls one can make to the API and the fact that with the random ID assignment, the number of invalid ID strings is tiny compared to the actual number of possible IDs. In fact, there are 1024 possible IDs for each millisecond, and on average, between zero to two accounts are valid (we found it from testing). Hence, a random user ID seldom returns a match. Therefore, we did not include this method in this paper.

\section{Robustness Tests}
\subsection{Regression Settings: Including District of Columbia}

\begin{figure}[H]
    \centering
    \footnotesize
    \includegraphics[scale=0.15]{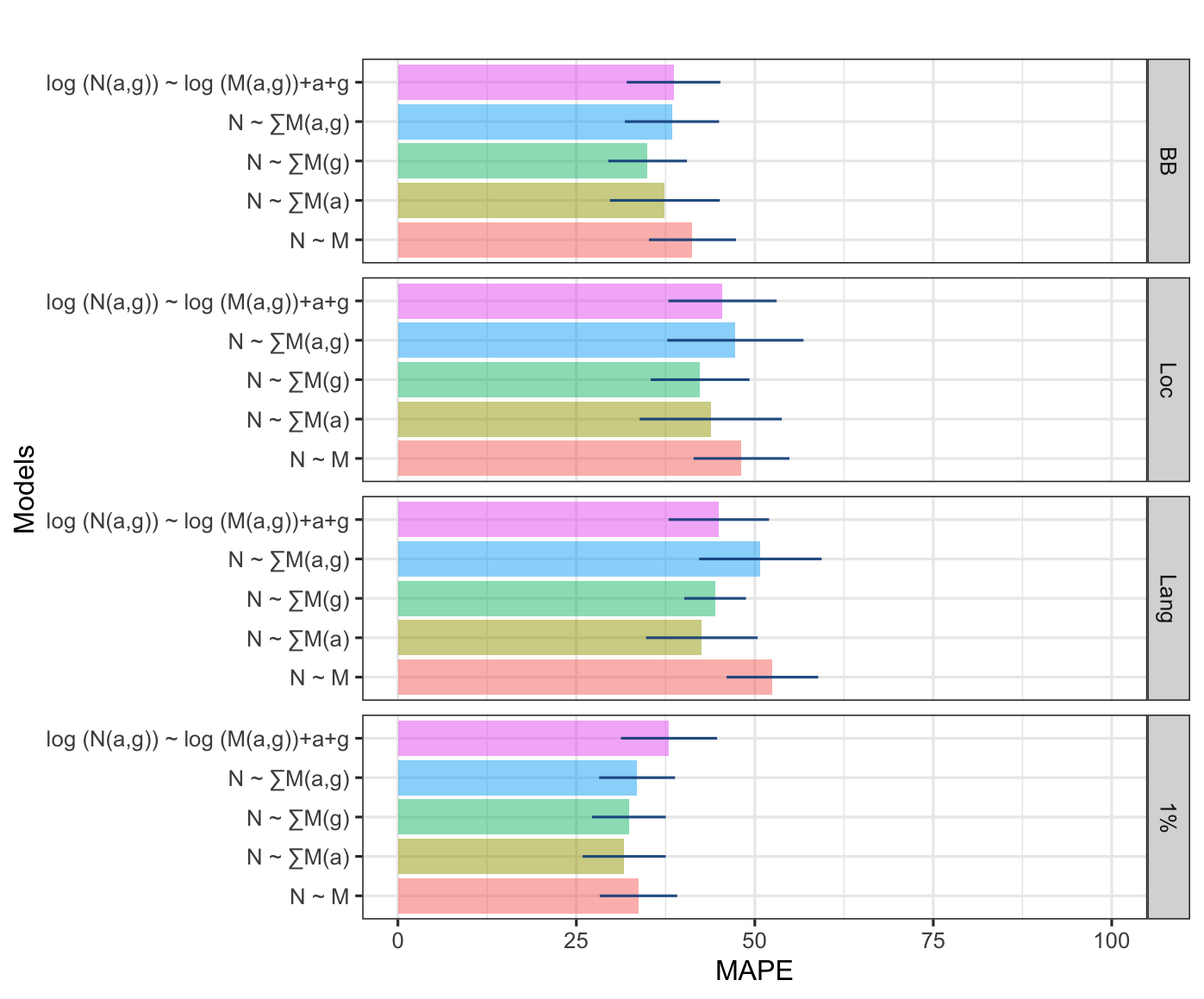}
    \caption{Performance on leave one state out population inference across different debiasing models, where the District of Columbia is included and rows with all zero value were removed from the regression. The bar shows MAPE($N$) robust standard errors clustered on states.}
    \label{fig:mape-51}
\end{figure}

\subsection{Regression Settings: Removing All States With At Least One Missing Group}

\begin{figure}[H]
    \centering
    \footnotesize
    \includegraphics[scale=0.15]{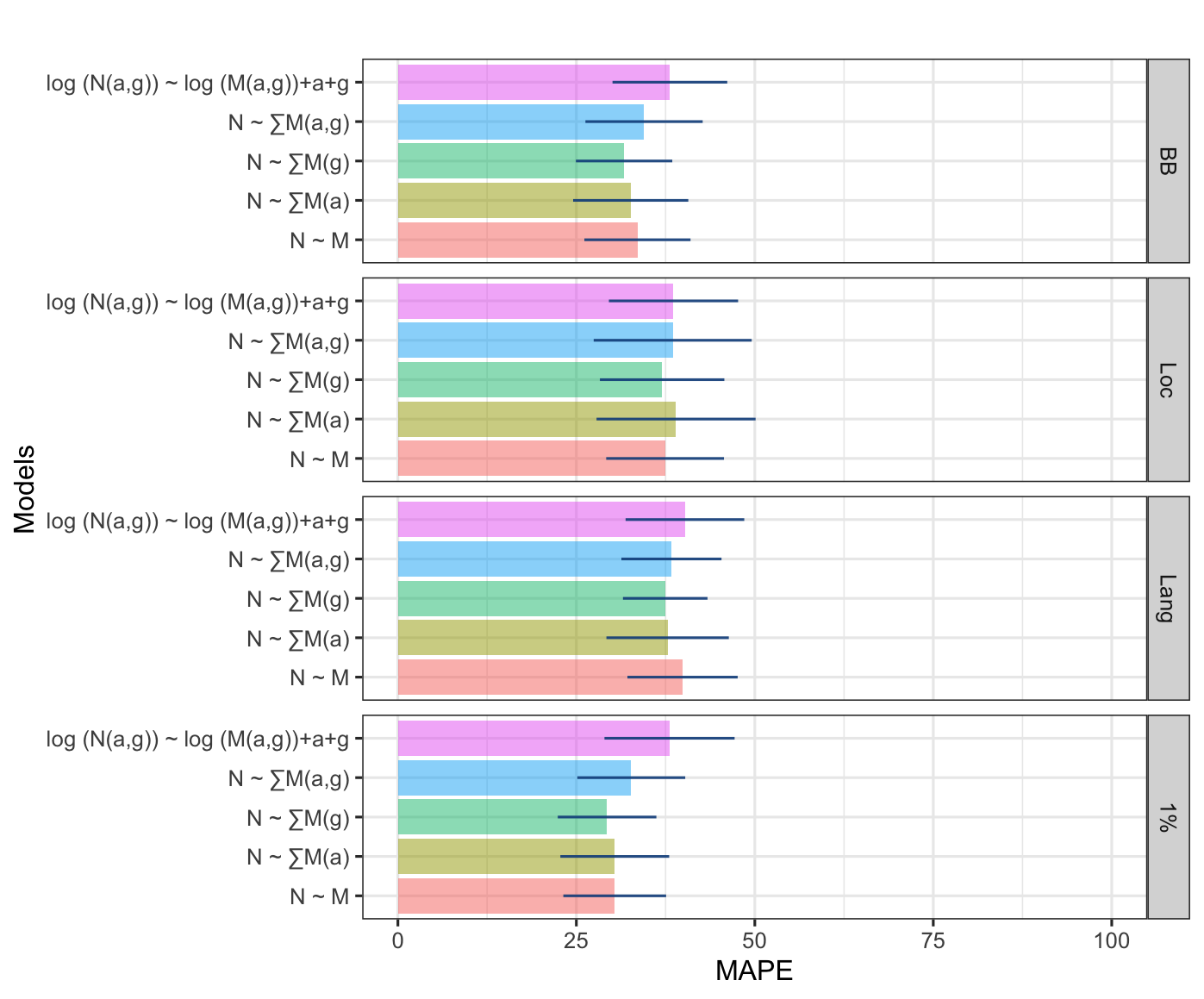}
    \caption{Performance on leave one state out population inference across different debiasing models, where all states with at least one missing demographic group in their Twitter data is removed from the regression. The bar shows MAPE($N$) robust standard errors clustered on states.}
    \label{fig:mape-39}
\end{figure}

\subsection{Regression Settings: Leave-One-Division-Out}

\begin{figure}[H]
    \centering
    \footnotesize
    \includegraphics[scale=0.15]{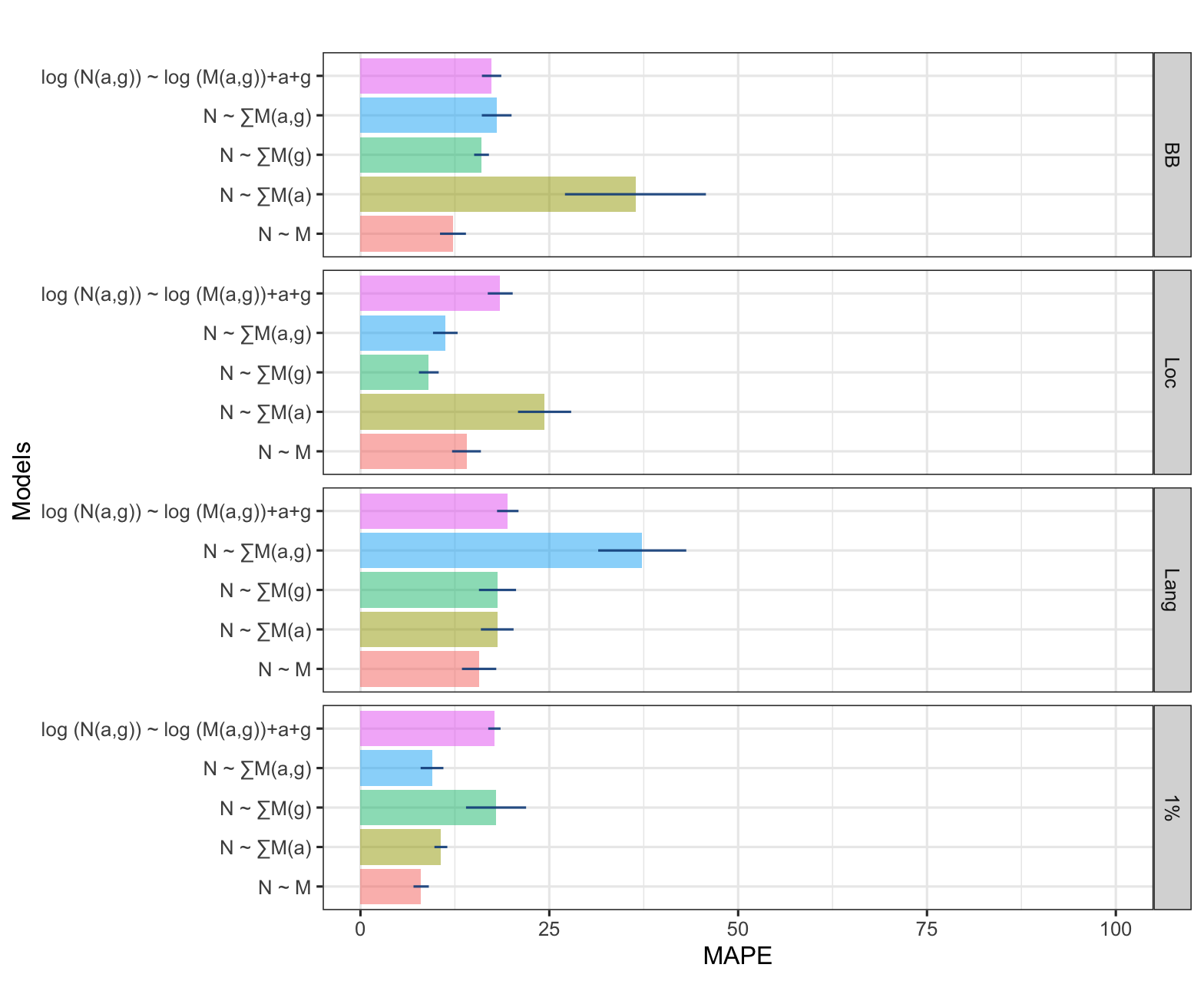}
    \caption{Performance on leave one division out population inference across different debiasing models. The bar shows MAPE($N$) robust standard errors clustered on US divisions.}
    \label{fig:mape-div}
\end{figure}

\subsection{Pre-Processing Settings: Removing Users with Less than 200 Tweets}

\begin{figure}[H]
    \centering
    \footnotesize
    \includegraphics[scale=0.15]{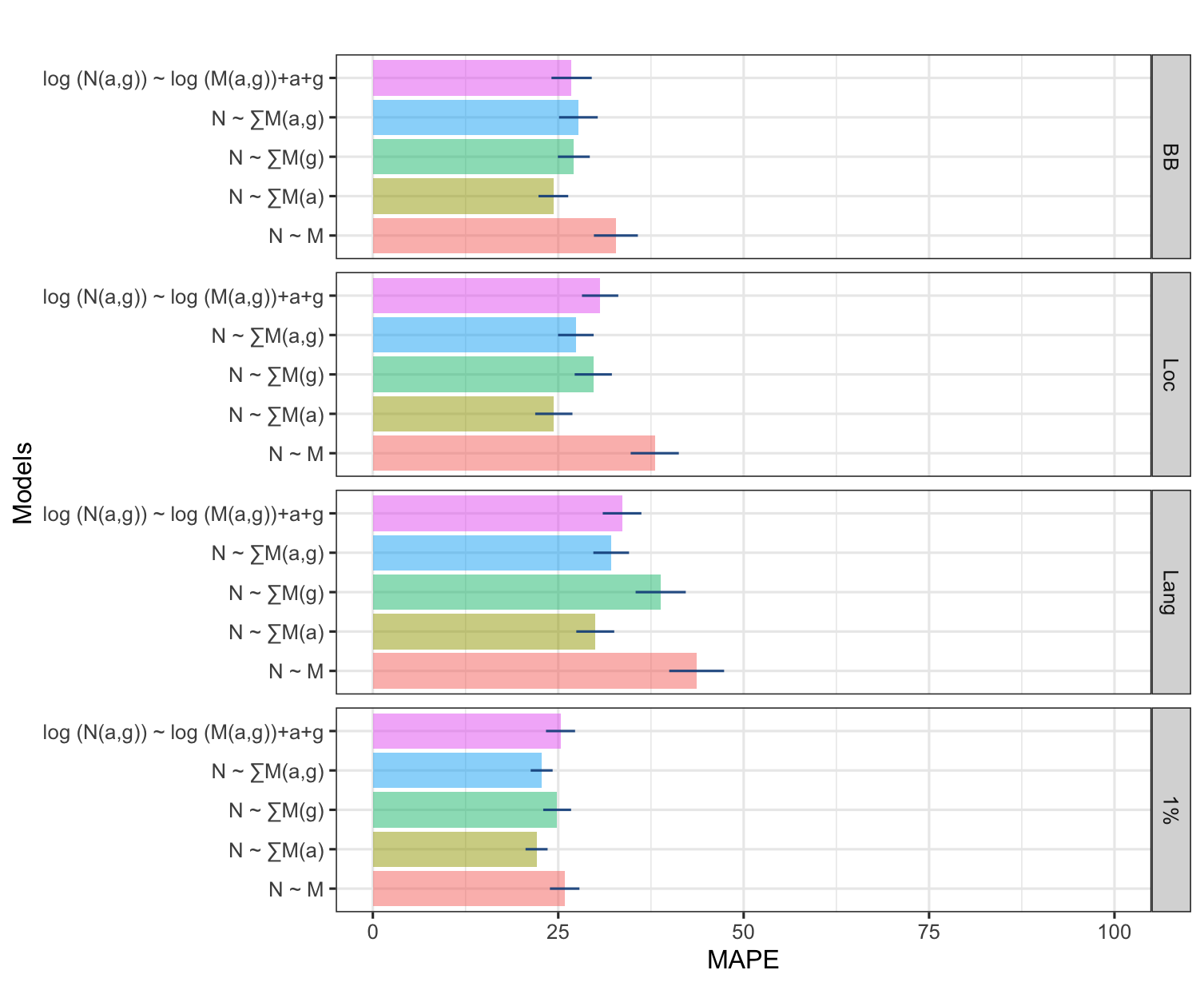}
    \caption{Performance on leave one state out population inference across different debiasing models where rows with all zero value were removed from the regression and minimum number of tweets filter changed from 100 to 200. The bar shows MAPE($N$) robust standard errors clustered on states. }
    \label{fig:mape200}
\end{figure}

\begin{figure}[H]
    \centering
    \begin{subfigure}{0.2\textwidth}
        \caption{}
        \label{fig:tweet_dist4}
        \includegraphics[scale=0.25]{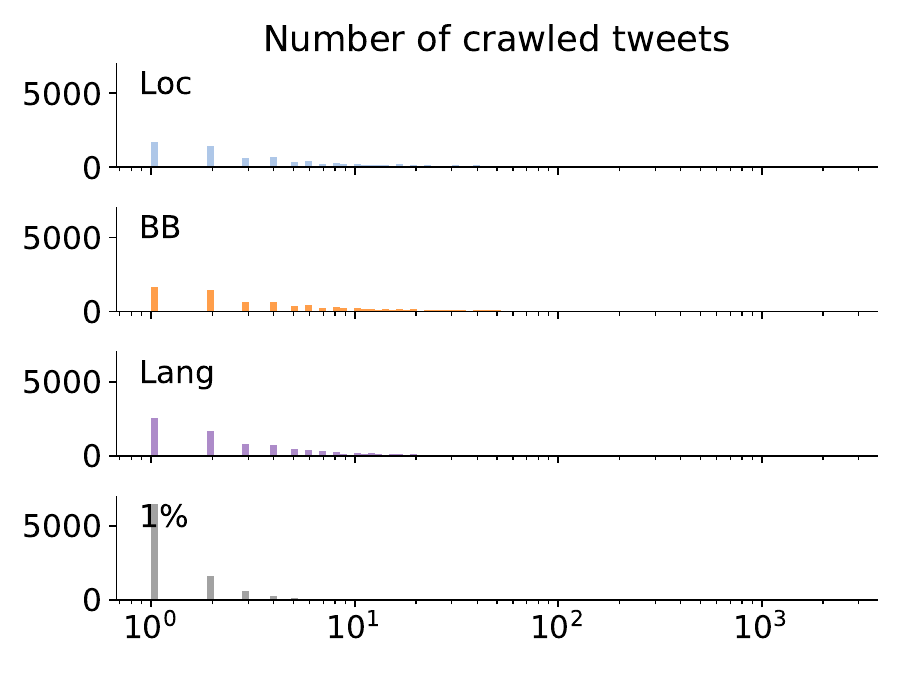}
    \end{subfigure}
    \begin{subfigure}{0.2\textwidth}
        \caption{}
        \label{fig:tweet_per_day4}
        \includegraphics[scale=0.25]{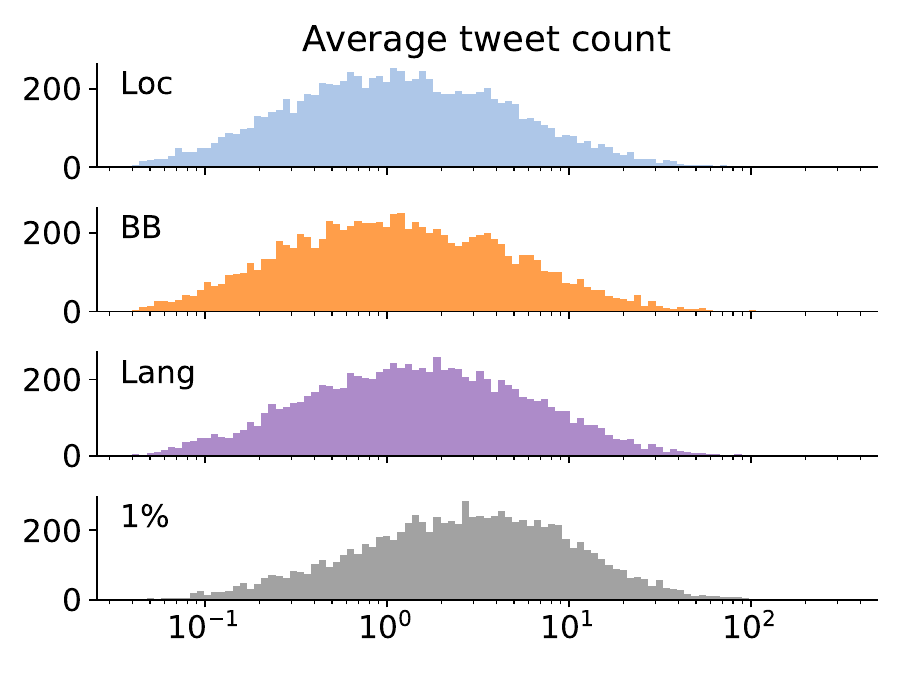}
    \end{subfigure}
    \\
    \begin{subfigure}{0.2\textwidth}
        \caption{}
        \label{fig:like4}
        \includegraphics[scale=0.25]{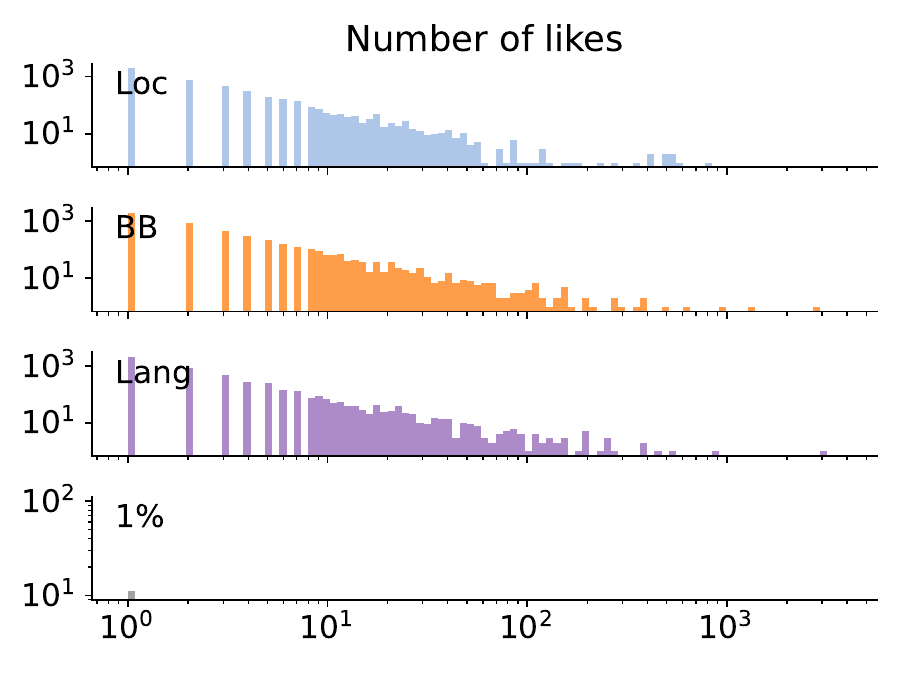}
    \end{subfigure}
    \begin{subfigure}{0.2\textwidth}
    \caption{}
        \label{fig:tenure4}
        \includegraphics[scale=0.25]{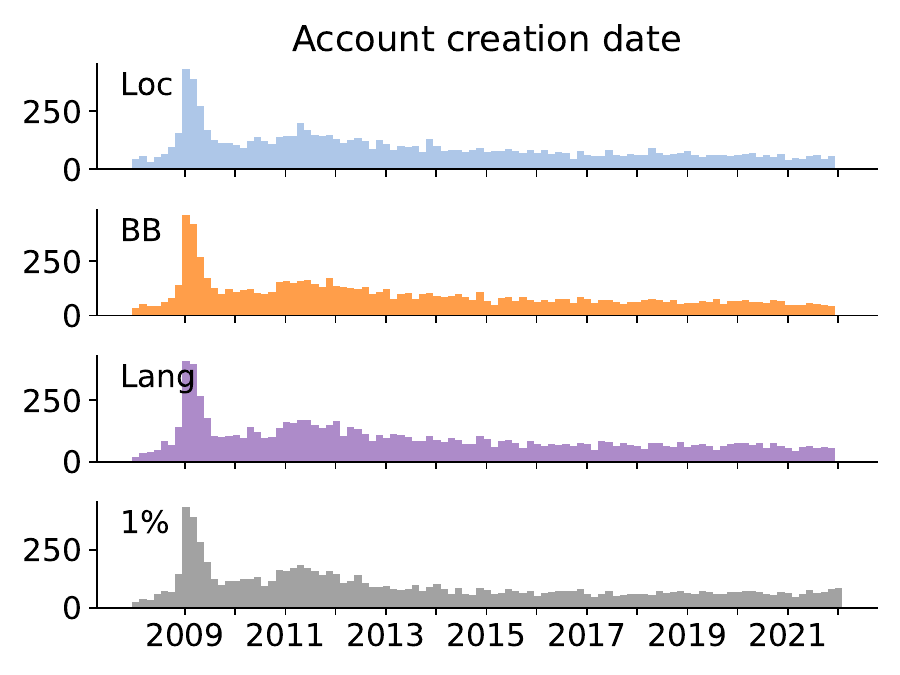}
    \end{subfigure}
    \\
        \begin{subfigure}{0.2\textwidth}
        \caption{}
        \label{fig:follower4}
        \includegraphics[scale=0.25]{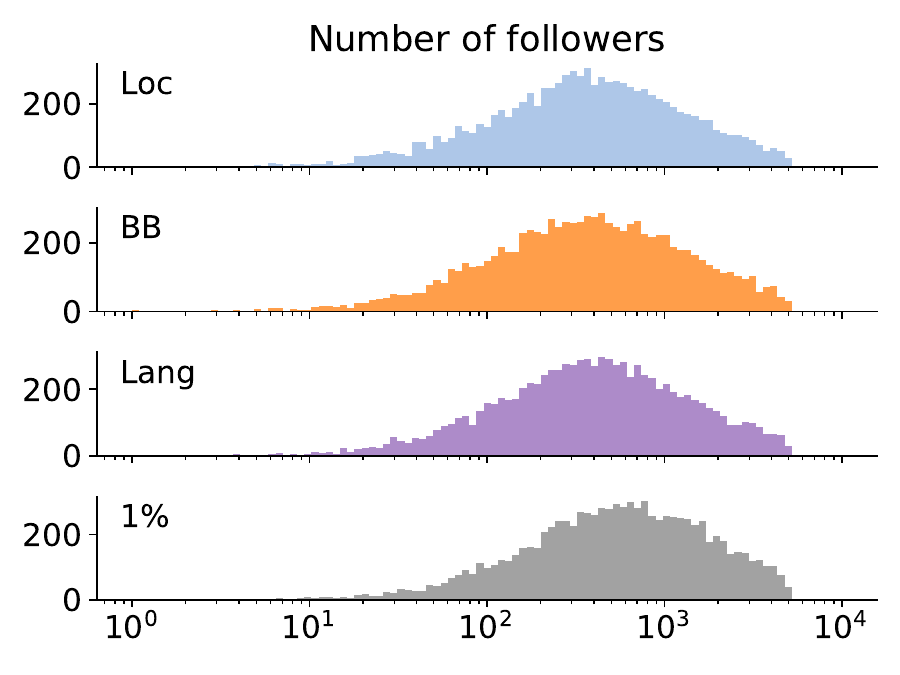}
    \end{subfigure}
    \begin{subfigure}{0.2\textwidth}
        \caption{}
        \label{fig:following4}
        \includegraphics[scale=0.25]{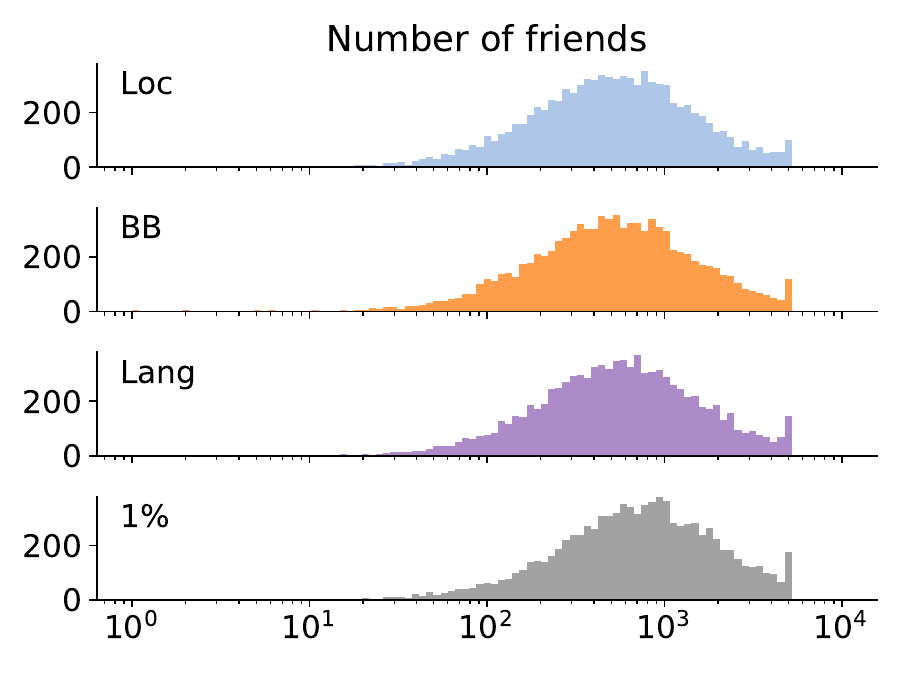}
    \end{subfigure}
    \\
    \begin{subfigure}{0.2\textwidth}
        \caption{}
        \label{fig:gender4}
        \includegraphics[scale=0.25]{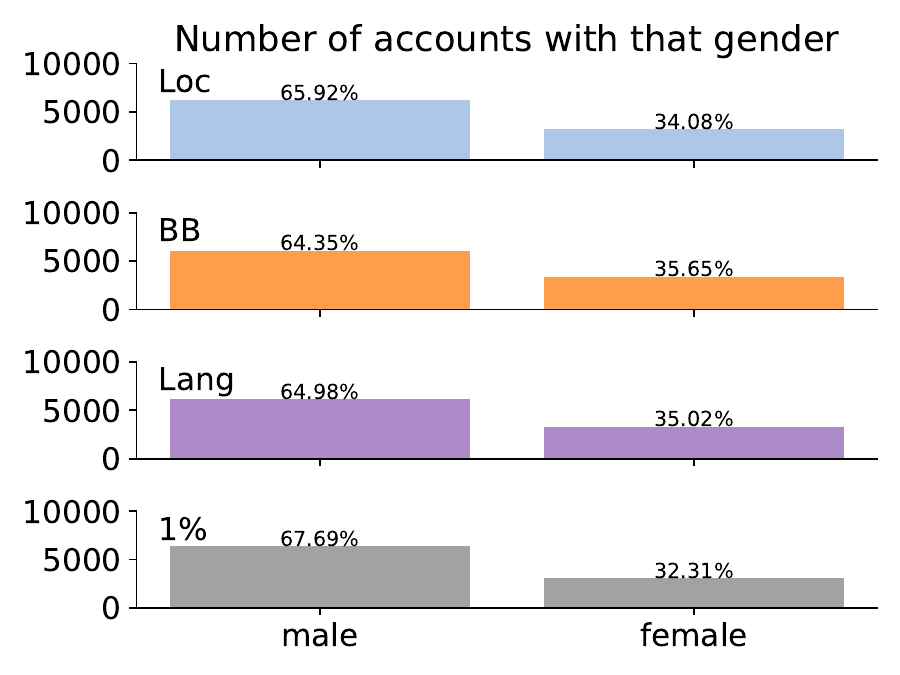}
    \end{subfigure}
    \begin{subfigure}{0.2\textwidth}
        \caption{}
        \label{fig:age4}
        \includegraphics[scale=0.25]{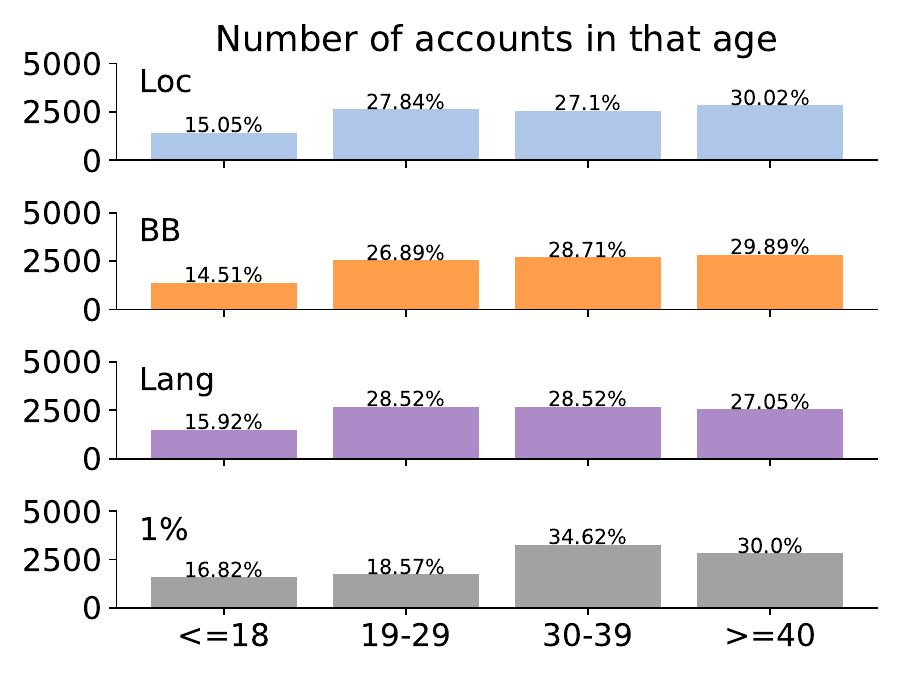}
    \end{subfigure} 
\caption{Distributions of (A) number of tweets; (B) average number of tweets per day; (C) number of likes; (D) account creation date; (E) number of followers; and (F) number of friends for different groups. Distribution of users with respect to (G) gender and (H) age across the four Twitter sampling methods after changing the minimum number of tweets filter from 100 to 200. After removing accounts with more than 200 tweets, the sample size decreased to 9,437. Consequently, we randomly selected an equivalent sample size for the remaining three samples.}
\label{fig:accountcompare200}
\end{figure}

\subsection{Pre-Processing Settings: Removing Users with Less than 1 Year Account Age}

\begin{figure}[H]
    \centering
    \begin{subfigure}{0.2\textwidth}
        \caption{}
        \label{fig:tweet_dist12}
        \includegraphics[scale=0.29]{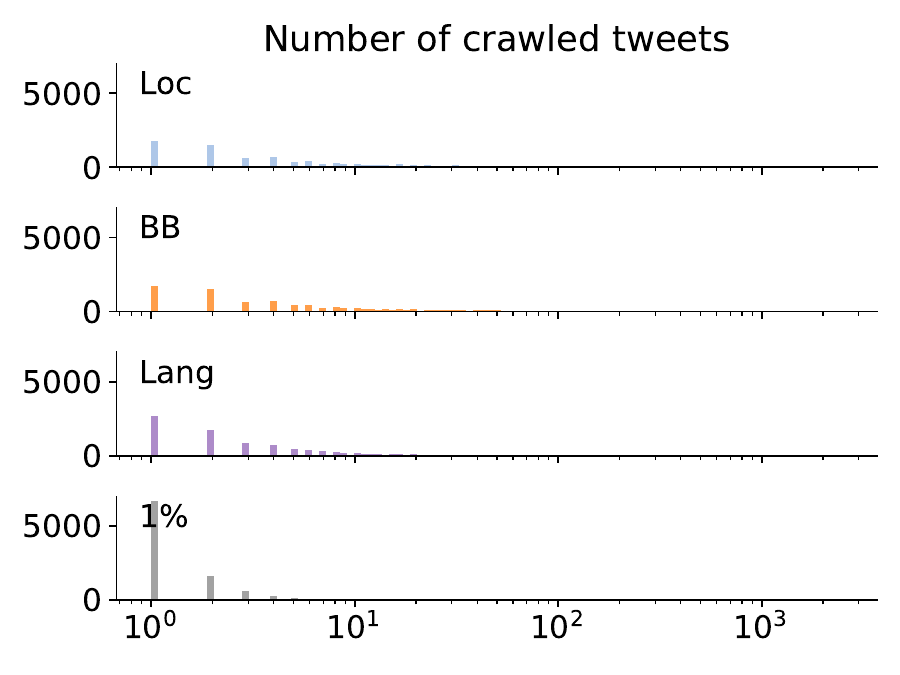}
    \end{subfigure}
    \begin{subfigure}{0.2\textwidth}
        \caption{}
        \label{fig:tweet_per_day12}
        \includegraphics[scale=0.29]{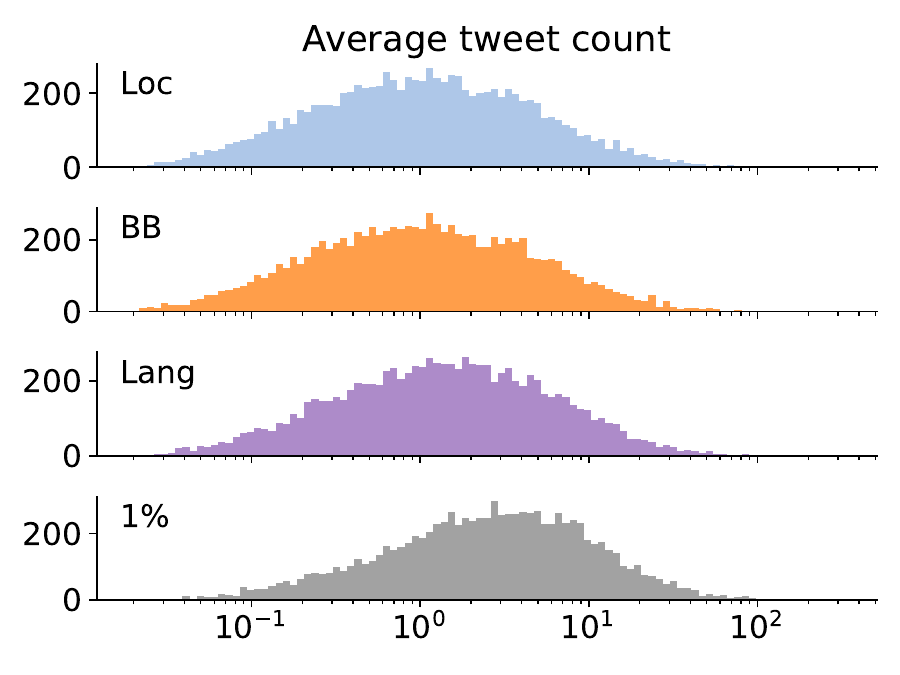}
    \end{subfigure}
    \\
    \begin{subfigure}{0.2\textwidth}
        \caption{}
        \label{fig:like12}
        \includegraphics[scale=0.29]{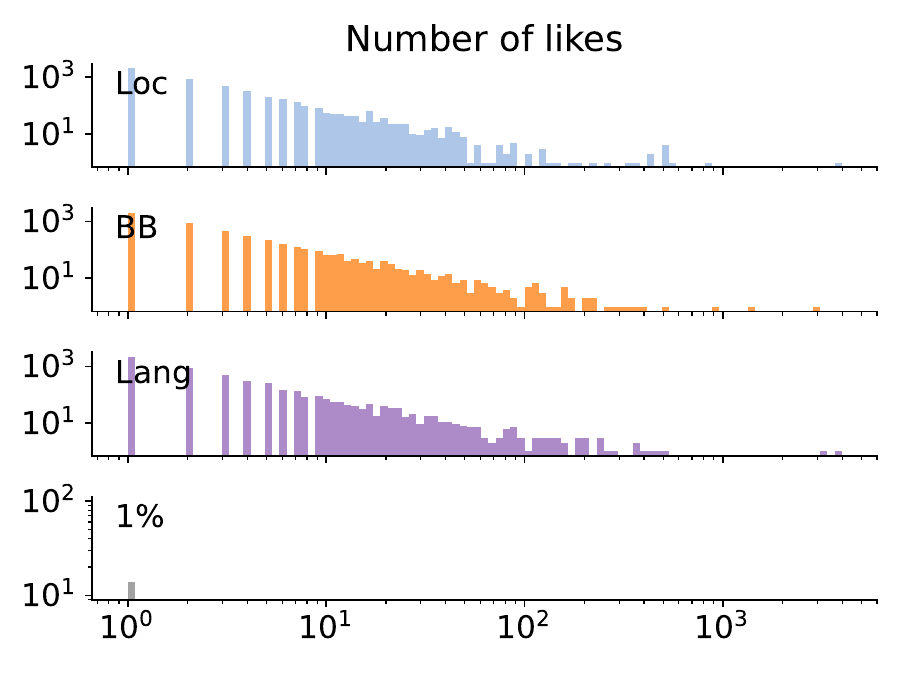}
    \end{subfigure}
    \begin{subfigure}{0.2\textwidth}
    \caption{}
        \label{fig:tenure12}
        \includegraphics[scale=0.29]{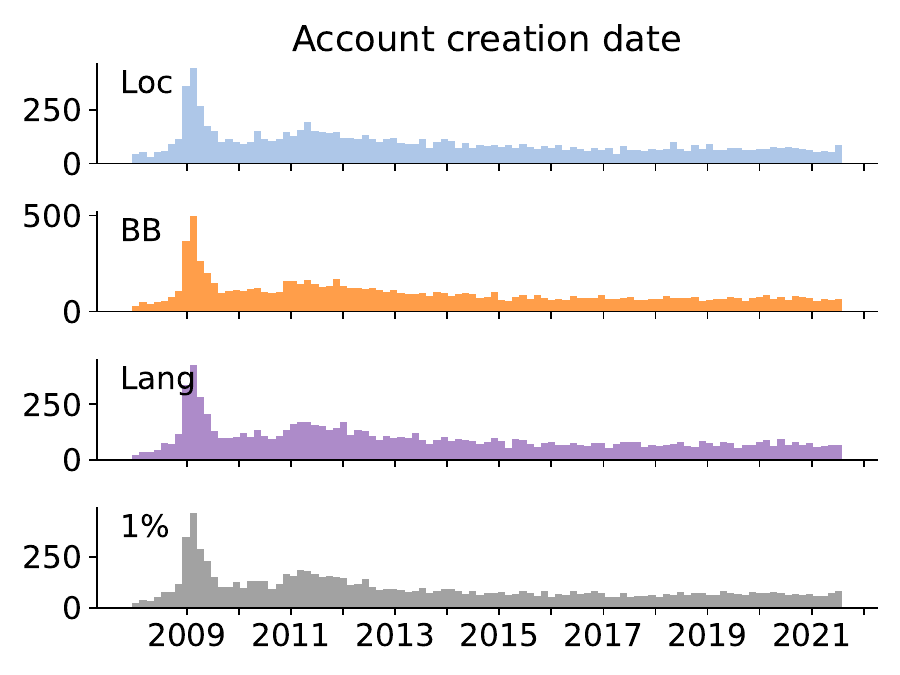}
    \end{subfigure}
    \\
    \begin{subfigure}{0.2\textwidth}
        \caption{}
        \label{fig:follower12}
        \includegraphics[scale=0.29]{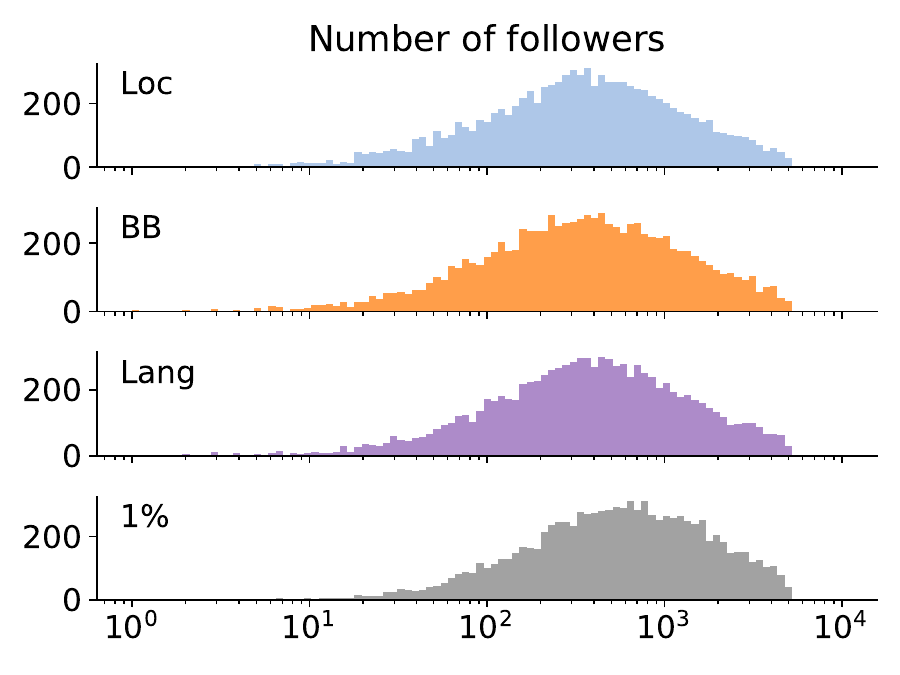}
    \end{subfigure}
    \begin{subfigure}{0.2\textwidth}
        \caption{}
        \label{fig:following12}
        \includegraphics[scale=0.29]{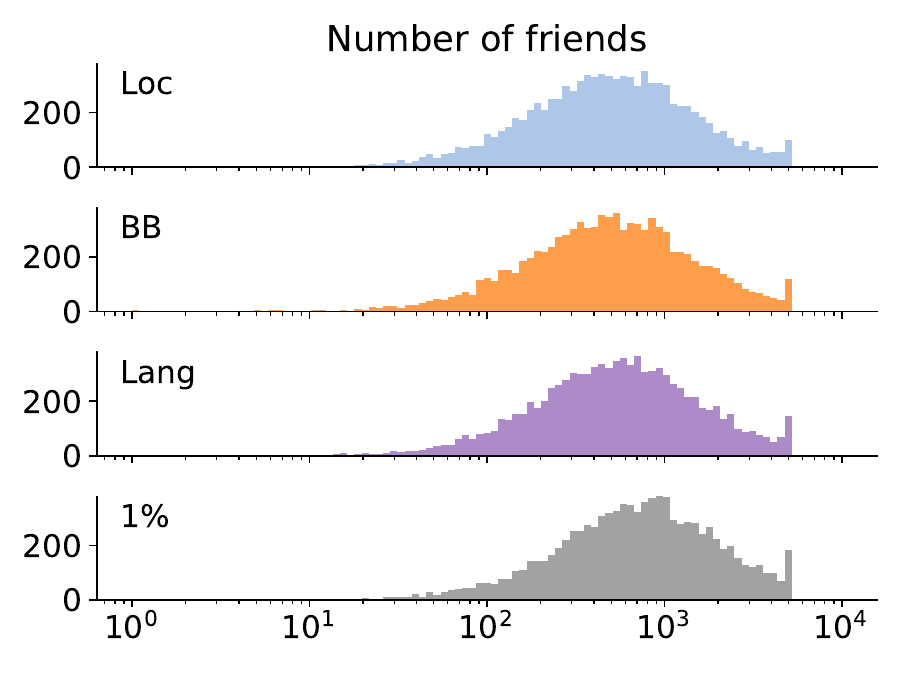}
    \end{subfigure}
    \\
    \begin{subfigure}{0.2\textwidth}
        \caption{}
        \label{fig:gender12}
        \includegraphics[scale=0.29]{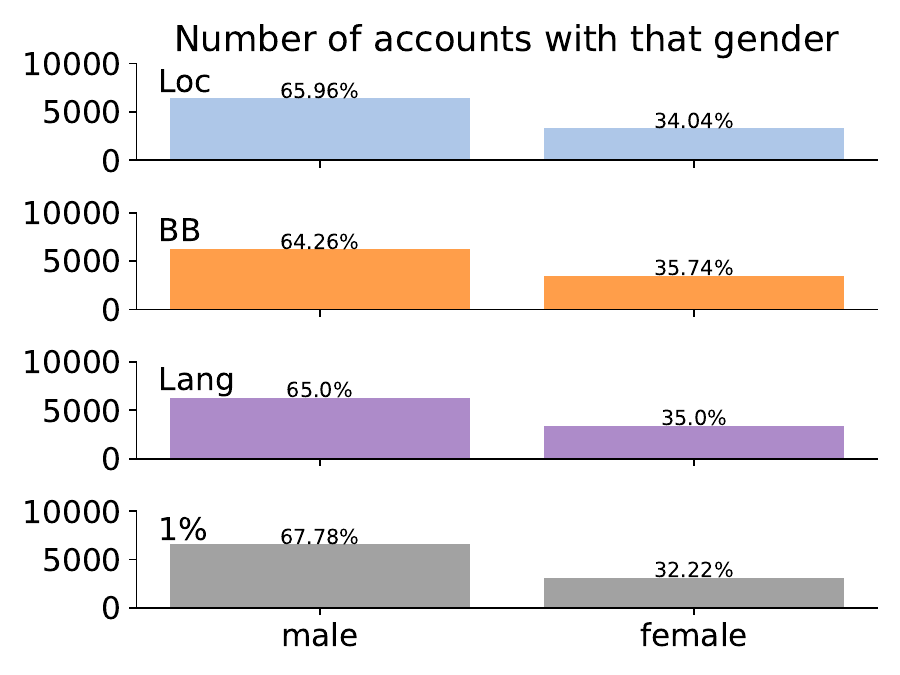}
    \end{subfigure}
    \begin{subfigure}{0.2\textwidth}
        \caption{}
        \label{fig:age12}
        \includegraphics[scale=0.29]{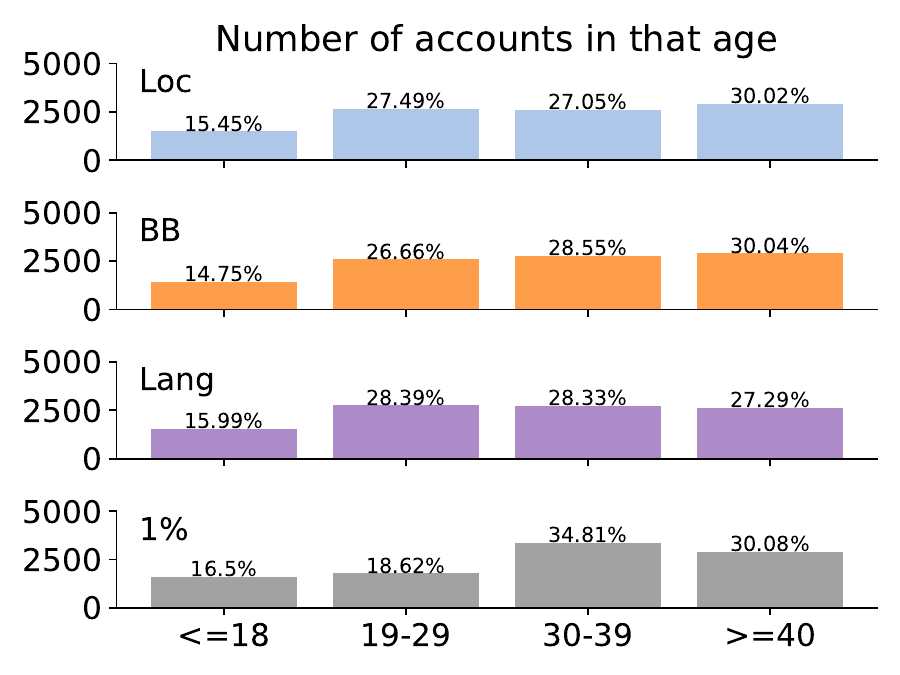}
    \end{subfigure} 
\caption{Distributions of (A) number of tweets; (B) average number of tweets per day; (C) number of likes; (D) account creation date; (E) number of followers; and (F) number of friends for different groups. Distribution of users with respect to (G) gender and (H) age across the four Twitter sampling methods after changing the minimum age of accounts filter from 9 months to one year. After excluding accounts created within the past year, the sample size was further reduced to 9,691. Subsequently, we matched this sample size for the remaining three samples.}
\label{fig:accountcompare12}
\end{figure}

\begin{figure}[H]
    \centering
    \footnotesize
    \includegraphics[scale=0.15]{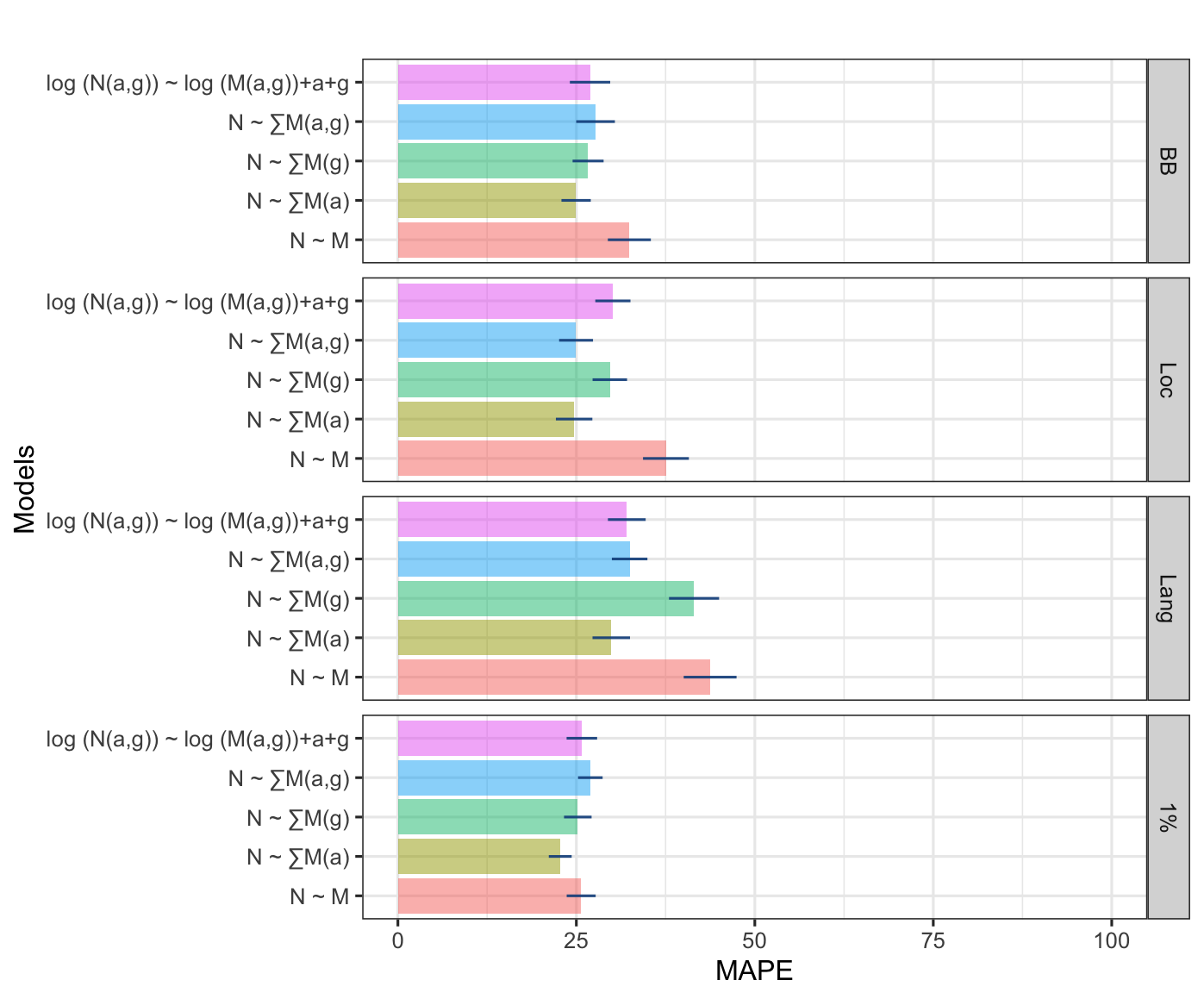}
    \caption{Performance on leave one state out population inference across different debiasing models where rows with all zero value were removed from the regression and the minimum age of accounts filter changed from 9 months to one year. The bar shows MAPE($N$) robust standard errors clustered on states.}
    \label{fig:mape12}
\end{figure}

\end{document}